\documentclass[5p,authoryear,times]{elsarticle}

\usepackage{amsmath}
\usepackage{amssymb}
\usepackage{txfonts}
\usepackage{epsfig,graphicx}

\usepackage{enumitem}

\usepackage[colorlinks=true, linkcolor=blue, citecolor=blue, urlcolor=blue]{hyperref}

\bibpunct{(}{)}{;}{a}{}{,} 

\journal{Astronomy and Computing}

\DeclareTextFontCommand{\mytexttt}{\ttfamily\hyphenchar\font=45\relax}

\makeatletter
\def\@author#1{\g@addto@macro\elsauthors{\normalsize%
    \def\baselinestretch{1}%
    \upshape\authorsep#1\unskip\textsuperscript{%
      \ifx\@fnmark\@empty\else\unskip\sep\@fnmark\let\sep=,\fi
      \ifx\@corref\@empty\else\unskip\sep\@corref\let\sep=,\fi
      }%
    \def\authorsep{\unskip,\space}%
    \global\let\@fnmark\@empty
    \global\let\@corref\@empty  
    \global\let\sep\@empty}%
    \@eadauthor={#1}
}
\makeatother

\begin{document}

\begin{frontmatter}

\title{SKIRT: hybrid parallelization of radiative transfer simulations}

\author{S. Verstocken\corref{cor1}}
\cortext[cor1]{Corresponding author. E-mail address: sam.verstocken@ugent.be}

\author{D. Van De Putte}
\author{P. Camps}
\author{M. Baes}

\address{Sterrenkundig Observatorium Universiteit Gent, Krijgslaan 281 S9, B-9000 Gent, Belgium}

\begin{abstract}
We describe the design, implementation and performance of the new hybrid parallelization scheme in our Monte Carlo radiative transfer code SKIRT, which has been used extensively for modeling the continuum radiation of dusty astrophysical systems including late-type galaxies and dusty tori. The hybrid scheme combines distributed memory parallelization, using the standard Message Passing Interface (MPI) to communicate between processes, and shared memory parallelization, providing multiple execution threads within each process to avoid duplication of data structures. The synchronization between multiple threads is accomplished through atomic operations without high-level locking (also called lock-free programming). This improves the scaling behavior of the code and substantially simplifies the implementation of the hybrid scheme. The result is an extremely flexible solution that adjusts to the number of available nodes, processors and memory, and consequently performs well on a wide variety of computing architectures.
\end{abstract}




\end{frontmatter}


\section{Introduction}
\label{intro.sec}

\noindent
In the study of astrophysical objects, the process of radiative transport plays a key role. One of the most challenging aspects of the study of radiative transfer is to incorporate the effects of dust. The radiative transfer problem is highly nonlocal in position, wavelength and direction space due to the scattering and reprocessing of optical and UV photons by the dust grains. This means that analytically, the radiative transfer equation would have to be solved simultaneously for all positions, directions and wavelengths. Moreover, the equation is also highly nonlinear, which makes it unfeasible to be solved except for very basic, low dimensional models.



\vspace{0.5em}
Fortunately, different methods have been conceived that make the problem numerically solvable for arbitrary 3D geometries. The now most commonly used technique, the Monte Carlo (MC) method, was introduced in dust radiative transfer around the 1970s \citep[e.g.][]{1970A&A.....9...53M, 1974ApJ...190...67R, 1974AJ.....79..948W, 1977ApJS...35....1W, 1977ApJS...35...31W}. Over the past decennia, the technique has been significantly refined and improved with additional acceleration mechanisms, including the peel-off method \citep{1984ApJ...278..186Y}, continuous absorption \citep{1999A&A...344..282L, 2003A&A...399..703N}, forced scattering \citep{ZAMM:ZAMM19600400727}, instantaneous dust emission \citep{2001ApJ...554..615B, 2005NewA...10..523B, 2006A&A...459..797P}, the library mechanism for dust emissivities \citep{2003A&A...397..201J, 2011ApJS..196...22B}, the diffusion approximation for high optical depths \citep{2009A&A...497..155M, 2010A&A...520A..70R} and variants of the photon packet splitting technique \citep{ZAMM:ZAMM19600400727, Jonsson2006, 2015MNRAS.448.3156H}. All of these mechanisms have enabled us to study the effects of dust on the radiation of a wide variety of systems, such as clumpy dust tori around active galactic nuclei (AGN) \citep{2012MNRAS.420.2756S}, circumstellar disks \citep{2006A&A...459..797P, VidalBaes2007}, Young Stellar Objects (YSO) \citep[e.g.][]{Whitney2013}, molecular clouds \citep{Boissel1990} etc. For a full overview of three-dimensional dust radiative transfer codes and their main applications, see \citet{WHITNEY2011} and \citet{2013ARA&A..51...63S}.

\vspace{0.5em}
As the applications of dust radiative transfer have been progressing towards higher-resolution models and tackling more complex physical phenomena, the performance of radiative transfer codes has become increasingly important. Despite the various acceleration mechanisms developed for Monte Carlo radiative transfer (MCRT), the ever-increasing demand for computing resources of radiative transfer applications require the exploitation of highly parallel systems (such as high-performance computing clusters). 



\vspace{0.5em}
MCRT codes are inherently different from many other codes used in numerical astrophysics, including hydrodynamical codes. These codes lend themselves to massive parallelization since each processing core can be assigned to one particular part of the computational domain, for the entire course of the program. Communications between processing cores are generally nearest-neighbour, and global communications are almost never required. In radiative transfer, the problem is generally nonlocal in space, except in some approximate solution methods. Therefore, a completely different approach is required. The MCRT algorithm works by launching numerous photon packages with a certain wavelength and luminosity through the dusty system and determining random interaction positions. The effects of each individual photon package are therefore dispersed throughout the entire system, with the location of the absorption and scattering events being effectively unpredictable. Equivalently, following a specific photon package requires information about the dust distribution over the complete physical domain.

\vspace{0.5em}
Furthermore, the different wavelengths in a MCRT simulation are tightly coupled due to the presence of dust. To calculate the emission spectrum of a dust grain, absorption information is required from all wavelengths. It is therefore impossible to perform the simulation for each wavelength individually, eliminating yet another tempting parallelization strategy.

\vspace{0.5em}
Another factor that affects the parallelization of MCRT codes is the memory usage. These codes are generally \emph{memory-bound}, which means that the time spent in memory accesses and writes (which correspond to unused CPU cycles) is comparable to or greater than the time spent in actual computation (used CPU cycles). Due to the Monte Carlo sampling, memory access patterns are random, which adds to overhead induced by caching and synchronization. Applications that perform heavy computations on a small set of data, on the other hand, are \emph{CPU-bound}. The locality between subsequent memory accesses leads to less overhead.

\vspace{0.5em}




\vspace{0.5em}
In the past few years, several teams have adopted parallelization in their radiative transfer codes. The HYPERION code \citep{Robitaille2011} features distributed-memory parallelization with MPI and achieves a speedup of 1604 (i.e. an efficiency of 0.2) on 8192 cores. 
In the SUNRISE code \citep{Jonsson2006}, shared-memory parallelization has been adopted using the PThread library. Another code, TORUS \citep{2015MNRAS.448.3156H}, which combines (aspects of) radiative transfer and hydrodynamics, uses a hybrid scheme with shared-memory and distributed memory programming. The distributed memory parallelization in TORUS relies on MPI and the shared-memory parallelization on OpenMP. With this hybrid method, the TORUS code was found to scale well up to (at least) 60 cores. \citet{HeymannSiebenmorgen2012} describe a MCRT code which, besides shared-memory parallelization based on OpenMP, uses yet another form of parallelization that enables the MCRT algorithm to run on graphics processing units (GPUs).

\vspace{0.5em}
In this work, we present a novel approach to the problem of parallelization in radiative transfer codes that is efficient and flexible enough to solve most of the problems described above. 
It is based on a task-based parallelization paradigm that distributes work elegantly among the cores of a parallel machine or across multiple nodes of a computing cluster. In cases where it is more efficient, it can also split up the main data structures used in the RT algorithm across nodes, substantially increasing the possible size of the problems that can be solved and making optimum use of the computing resources available. The scheme incorporates so-called \emph{lock-free} operations at critical points, a mechanism that may not be widely known within the astrophysical community but nevertheless can have an substantial impact on performance.

\vspace{0.5em}
The next section will introduce and define some parallelization concepts, including a discussion of lock-free programming. Section \ref{skirt.sec} briefly presents SKIRT, and its past and current applications. In section \ref{modes.sec}, we elaborate on the different parallelization modes offered by the new version of SKIRT, motivated by the variety of use cases for which SKIRT is intended. 
In section \ref{design_and_implementation.sec}, the design and implementation of our parallelization approach will be discussed. Section \ref{skirtwithparallelization.sec} contains instructions on how to use the parallelization. In section \ref{results.sec}, we present the results of a series of tests that were designed to gauge the performance and efficiency of the implementation. We also show the output of a high-resolution simulation generated with the new parallelized version of SKIRT. Conclusions are given in section \ref{conclusions.sec}.

\vspace{0.5em}
This work fits in a series about the radiative transfer code SKIRT, of which the features and interface are introduced by \citet{CampsBaes2015} and the design of input models by \citet{BaesCamps2015}.

\section{Parallelization concepts}
\label{parallelization.sec}

\subsection{Shared and distributed memory}
\label{shareddistributed.ssec}

\vspace{0.5em}
\noindent
In a multicore computing system, processors and memory can be arranged in a multitude of ways. Whereas in typical personal workstations and laptops processing cores are connected to the same unit of memory through the motherboard, larger computing systems are often arranged as a network of individual \emph{nodes}. The nodes have their own chipset with processing, memory and IO modules. Each node of a computing cluster, just like a personal computer, acts as a so-called \emph{shared-memory system}. The cluster itself, as a combination of shared-memory systems, is called a \emph{distributed-memory system}. On such a system, memory access is non-uniform: a random pair of cores either shares memory or not. If they don't share memory, and one processor needs information stored in memory at another processor, explicit messages will have to be exchanged between their respective nodes via the network. 

\vspace{0.5em}
The difference between shared memory and distributed memory is the clearest distinction that can be made between memory-CPU topologies. The layout of memory and execution units within one node, however, can also be organized in many ways. `Shared' memory can be seen as a relative term in this context, since memory may be more local to one execution core than to another within the same node. 

\begin{figure}[tb]
\centering
\includegraphics[width=0.49\textwidth]{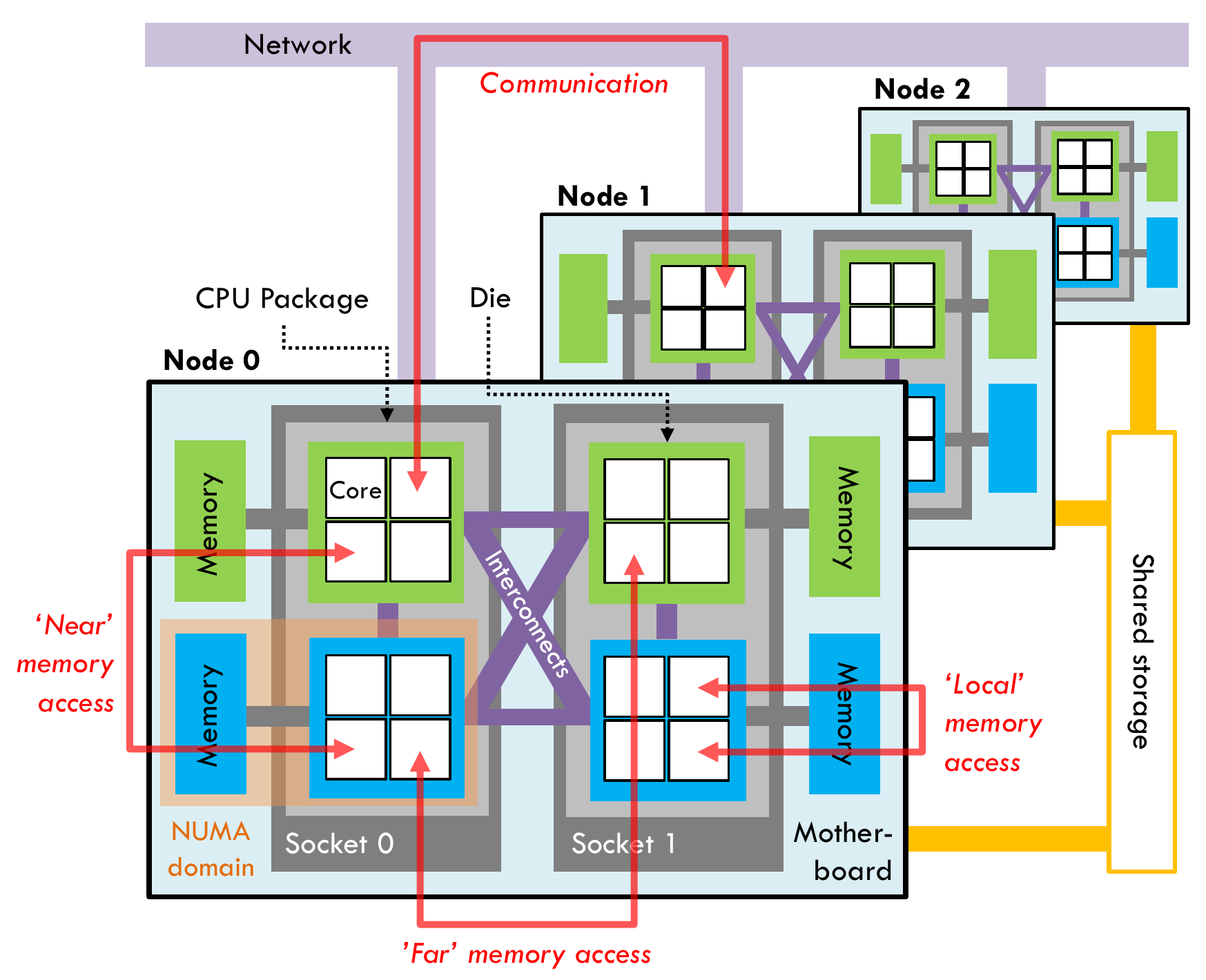}
\caption{Schematic overview of the topology of a computing cluster. In this configuration, each computation node contain two sockets, occupied by a CPU package consisting of two dice. Each die contains four cores, and the the two dice on the package are connected with a fast interconnection bridge. A separate memory module is connected to each of the dice, which allows direct access from each of its cores. In other words, information is exchanged between these cores by simply storing it in the same local memory module. Data exchange between cores on the same CPU package but on a different die is slower than local exchange because it uses the interconnection between the dice. Other interconnect paths are provided between the two sockets of the node, allowing `far' memory access between cores. Exchange of information between cores on different nodes requires messages to be sent over the node network.}
\label{cluster.fig}
\end{figure}

\vspace{0.5em}
Figure \ref{cluster.fig} shows a simplified diagram of the processing and memory layout of a hypothetical distributed-memory system. It shows the possible levels of compartmentalization of the logical and memory unit of such a system. The most basic level, used by the operating system for scheduling and executing sequences of instructions, is the logical core. Since the advent of \emph{simultaneous multithreading} \citep{Tullsen1995}, a physical core can perform more than one task at a time (increasing the throughput of computations) and thus represent more than one logical core. In Figure \ref{cluster.fig}, each physical core has been equipped with two \emph{hardware threads}. These hardware threads appear as individual logical cores to the operating system.

\vspace{0.5em}
Between node and (physical) core, there can be different other levels: cores are embedded in dice (plural of \emph{die}), which are combined into a CPU package, and a node can hold multiple packages, depending on the number of available CPU sockets. The socket on the motherboard provides the processor cores with links to the memory units (memory channels). While the CPU packages are part of the same shared-memory node, cores on the same package (or die) share memory that is \emph{more local} than that connected to the other package (or die). Memory access between cores on the same die and between cores on the same package can be called \emph{local} and \emph{near} memory access. Interconnections link the dice of different CPU packages to allow \emph{far memory access}. This design of memory topology within a shared-memory system is called \emph{non-uniform memory access} (NUMA) \citep{Bolosky1989}. Before NUMA, processors working under a single operating system (node) would access each other's memory over a common bus or interconnect path. The problem is that as more and more cores are added to this configuration, the contention for data access over the shared bus becomes a major  performance bottleneck. These symmetrical configurations of shared-memory systems are extremely difficult to scale past 8-12 CPU cores. NUMA reduces the number of cores competing for access to a shared memory bus by adding an intermediate level of processor-memory locality.



\vspace{0.5em}
Programming for a parallel machine requires a flexibility in strategies as diverse as the underlying architectures. This flexibility can be obtained by combining two parallelization paradigms, \emph{shared memory} parallelization (\emph{multithreading}) and \emph{distributed memory} parallelization (usually the \emph{message-passing} model).  

\vspace{0.5em}
On a shared-memory system (whether NUMA or not), it is beneficial if computations that run in parallel on multiple cores can share resources. This can be achieved by multi-threaded programming, which allows for multiple execution \emph{threads} to read from and write to the same data. Ideally, these data are then physically kept as locally as possible to the corresponding cores. Multithreading provides a favorable optimization method in Monte Carlo radiative transfer codes, since each of the tasks (calculating photon trajectories) requires the entire physical domain.




\vspace{0.5em}
The strength of multi-threaded parallelization, however, is also its weakness: since threads share memory, contentions for the same resource increase rapidly in frequency as more cores (threads) are added. Exclusively multi-threaded applications are therefore severely limited in terms of performance scaling. 

\vspace{0.5em}
When there is too much contention for the same resources or too much load on interconnecting memory buses, distributed memory parallelization provides a solution. With this model, resources are kept private to each execution unit (which is called a \emph{process} in this case). The advantage of this method is that since none of the memory is shared, there is no contention, and additionally, processes can allocate all their memory locally without affecting the load on interconnecting buses. The disadvantage is that data transfer between execution units requires sending explicit messages, often over a network across nodes. The strain on this network can be alleviated to some extent if a workable subdivision of the data can be achieved and processes can work on these subsets of data quasi-independently. The industry standard for this kind of parallelization is the Message Passing Interface (MPI) and the two dominant implementations are Open MPI \citep{Gabriel2004} and MPICH \citep{bridges1995user} (and their derivatives). The learning curve for the code developer is arguably steeper with MPI than with shared-memory solutions and the implementation in an existing algorithm is often non-trivial.


\vspace{0.5em}
In MCRT codes, the subdivision of data is not easily obtained, because of the nonlocality in all variables. 
\citet{2015MNRAS.448.3156H} however, describes a scheme where the spatial domain is divided into subdomains and photon packages are passed from one process to another when they cross a boundary between subdomains. To limit the resulting communication overhead, photon packages are stacked as serialized data blocks and only sent intermittently. This, inevitably, introduces a certain level of complexity to the code. 

\vspace{0.5em}
In this work, we argue that the nature of the MCRT algorithm lends itself more naturally to a task-based approach rather than a spatial domain decomposition. We present an algorithm that is decomposed in wavelength space in some parts, and decomposed in position space in other parts and demonstrate that the necessary shifting around of data is limited and efficient.

\vspace{0.5em}
We also show that a hybrid programming model, combining both shared-memory and distributed-memory parallelization, can alleviate much of the problems that each of these methods alone would face. The hybrid model is efficient because it can adapt to the specific architecture of the system. 

\vspace{0.5em}
Throughout this text, we will occasionally use the notations $N_{\text{t}}$ and $N_{\text{p}}$ to represent the number of threads and processes respectively.

\subsection{Lock-free programming}
\label{lockfreeprogramming.ssec}
\vspace{0.5em}

\noindent
As discussed in the previous sections, MCRT codes require numerous read and write procedures from and to memory during the trajectory of a photon package through the physical domain. In a multithreading environment, the write operations become troublesome when multiple threads concurrently try to change information in the same memory location.
Traditionally, \emph{locks} would have been implemented to prevent situations where two threads compete for the same resource and cause unexpected results (\emph{race conditions}). A lock is a special kind of a \emph{semaphore} \citep{DijkstraSeinpalen}, which is a more general construct used to control shared resources in concurrent environments.

\vspace{0.5em}
A lock acquired by a particular thread, prevents other threads in the application from accessing the contested resources. Thus, it can effectively prevent situations where outdated information is used by one thread to change the state of the program, erasing the contribution of another thread that had changed that information in the meantime.

\vspace{0.5em}
Locking, however, has its disadvantages. The clearest disadvantage is the runtime \emph{overhead} that is introduced. Each acquisition and subsequent release of a lock consumes some CPU time, introducing a certain overhead even when there is no contention for the corresponding resource at the time of locking. If there \emph{is} contention, one or more threads will be halted until the lock is released. Depending on the amount of locking that is required, the scalability of the application can be severely limited. Another disadvantage is that situations where none of the threads can make progress due to circular lock dependencies (so-called \emph{deadlocks}), have to be avoided by careful programming design.



\vspace{0.5em}
To avoid the downsides of locks, so-called \emph{lock-free} or \emph{non-blocking} algorithms were adopted. Lock-free algorithms use so-called \emph{atomic} operations to provide thread synchronization. An atomic operation, if supported by the underlying hardware\footnote{Atomic operations are supported by almost all current architectures. If not, the compiler adds an appropriate software lock.}, is translated into a single machine instruction. As a consequence, it is also indivisible: it either succeeds or has no effect, the intermediate state is never observed. This automatically avoids the race conditions discussed before. Although atomic operations avoid high-level locks, some form of locking is still used under the hood. It must be noted, however, that locking at the application level is explicit and coarse-grained (leading to definite overhead), while these locks, also called \emph{cache locks}, are implicit, fine-grained, low-cost, and interpreted at the lowest possible instruction level \citep{Herlihy1993, intel2012}.

\section{SKIRT radiative transfer}
\label{skirt.sec}

\subsection{General characteristics}
\label{generalskirt.ssec}
\vspace{0.5em}

\begin{figure*}[th!]
\centering
\includegraphics[width=0.85\textwidth]{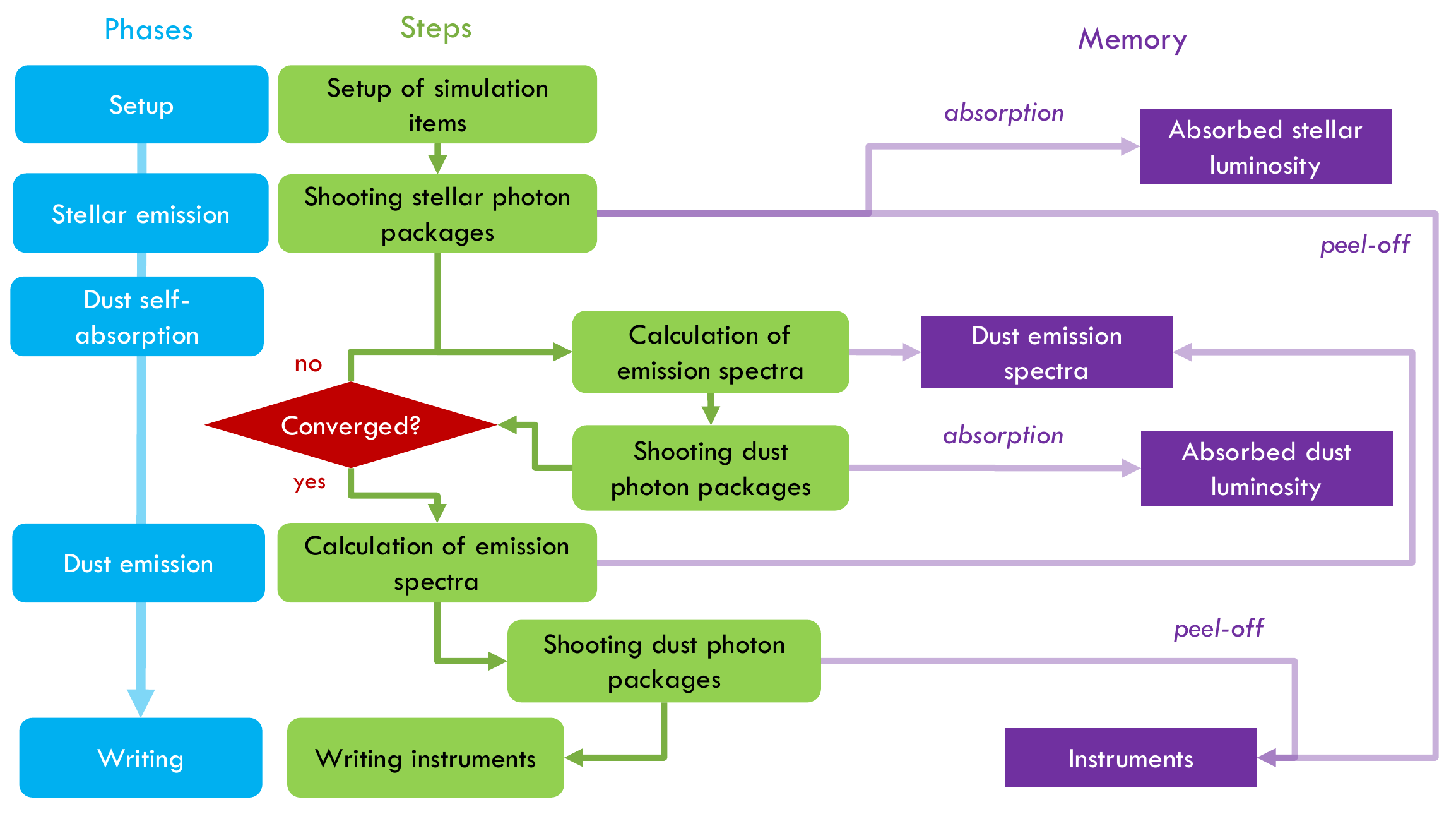}
\caption{A flowchart of the steps and data structures involved in a SKIRT simulation. To the left, in blue, the different simulation phases are shown. The second column, in green, lists the various steps corresponding to these phases and their order of execution. To the right, in purple, the most prominent data structures maintained by SKIRT are presented. Also indicated are the write procedures from the SKIRT algorithm to the data structures. Photon emission steps modify the absorbed luminosities in the dust system and in the instruments. The calculated dust emission spectra, re-calculated during each self-absorption step, are stored elsewhere.}
\label{skirtflow.fig}
\end{figure*}


\noindent
SKIRT \footnote{\hyperlink{www.skirt.ugent.be}{www.skirt.ugent.be}} is a state-of-the-art open-source 3D continuum MCRT code, written in C++ \citep{2011ApJS..196...22B, CampsBaes2015}. The source code is publicly available on GitHub \footnote{\hyperlink{https://github.com/SKIRT/SKIRT}{https://github.com/SKIRT/SKIRT}}. It provides a full treatment of scattering, absorption, emission by dust, and stochastic heating of small dust grains. It incorporates many of the optimization techniques discussed in the introduction. Being devised with highly modular design in mind, it provides an extensive variety of possible input geometries \citep{BaesCamps2015}, pre-defines many types of dust mixtures and supports advanced dust grids \citep{2013A&A...554A..10S, 2014A&A...561A..77S}. SKIRT also allows using the output of hydrodynamical simulations as input \citep{Saftly2015, CampsFIRpropEAGLE2016}, as well as observed FITS images that can be deprojected and turned into 3D geometries (\citet{2014A&A...571A..69D}, Verstocken et al., in prep.).


\vspace{0.5em}
SKIRT has been used to study the energy balance problem in edge-on spiral galaxies \citep{2010A&A...518L..39B, DeLooze2012, DeLooze2012b, Saftly2015, DeGeyter2015}, dust heating mechanisms in nearby galaxies \citep[Verstocken et al., in prep.]{2014A&A...571A..69D, VIAENE2016}, the observed properties of galaxies from cosmological hydro simulations \citep[Trayford et al., submitted]{CampsFIRpropEAGLE2016}, and the effects of a clumpy dust torus on AGN properties \citep{2012MNRAS.420.2756S, 2016MNRAS.458.2288S}. SKIRT has also been used to solve the so-called \emph{inverse} radiative transfer problem, where the 3D distribution of stars and dust in a particular galaxy is estimated based on UV, optical and NIR observations of that galaxy. For this purpose, a fully automatic fitting procedure FitSKIRT has been developed as an extension to SKIRT \citep{DeGeyter2013} and it has been applied to different galaxy types \citep[Mosenkov et al. in prep.]{2014MNRAS.441..869D, Viaene2015, 2016A&A...592A..71M}.


\subsection{Simulation phases}
\label{simulationphases.ssec}
\vspace{0.5em}

\noindent
Figure \ref{skirtflow.fig} gives an overview of the different steps that are traversed during a typical SKIRT radiative transfer simulation. On the left, the progress of the main simulation phases is depicted. As described by \citet{2011ApJS..196...22B}, these phases include the setup, the stellar emission phase, the dust self-absorption phase, the dust emission phase, and the writing phase. For each of these phases, there are a few substeps, represented as green boxes in the figure. The purple boxes illustrate the main data structures that are maintained in memory during a SKIRT simulation. These include the instruments, two data containers for the absorbed luminosities in the dust system, and a data container for the dust emission spectra. Each simulation step has a specific interaction with these data structures. Figure \ref{skirtflow.fig} only shows write operations to the memory.

\noindent
\paragraph{Setup} During the setup, the dust and stellar systems are initialized based on information stored in data files (bundled with SKIRT) and the information or choices provided by the user. The most important aspect of the setup is the construction of the dust grid, a discretization of the dust distribution defined by analytical functions or the output of hydrodynamical simulations. The entities of the dust grid are called dust cells.



\noindent
\paragraph{Stellar emission phase} During the stellar emission phase, the transfer of radiation produced by the primary sources is simulated. These sources are usually stellar in nature, but they can also be any other source of which the emission is fixed and known a priori (e.g. templates of the spectrum of star-forming regions). For each wavelength of the simulation, a fixed number of photon packages is propagated through the system. Monte Carlo sampling is used to determine the launch position, interaction positions and interaction directions. A technique called \emph{peel-off} is used to direct a fraction of the luminosity of a package towards the instruments on each scattering event, to improve the sampling of the emergent radiation field in the observer's field of view \citep{1984ApJ...278..186Y, 2011ApJS..196...22B}. SKIRT stores a table of the luminosity that is collected in each pixel of the instrument for each wavelength separately (the instrument table). A typical photon package will encounter multiple scattering events (depending on the optical depth) during its lifetime, until it has lost all but a neglegible part of its original luminosity. The arrow from the stellar photon shooting step towards the instruments in Figure \ref{skirtflow.fig} depicts the luminosity that is added to the instrument table during each scattering event. In between scattering events, a photon package also gradually loses luminosity through absorption by the dust grains. This is illustrated by the arrow pointing to the absorbed stellar luminosity table. This table stores the luminosity stored in each dust cell and for each wavelength in the simulation. 

\noindent
\paragraph{Dust self-absorption phase} When all stellar photons have been simulated, the absorbed luminosities in each dust cell are adjusted for absorption of photons emitted by the heated dust grains. This is performed in the dust self-absorption phase. First, the emission spectrum is calculated for each dust cell, based on the amount of absorbed stellar radiation. The dust emission spectra table stores the emission spectra for all dust cells. Next, photon packages are launched from the dust cells according to these emission luminosities. The mechanism of this photon shooting step is similar to the one in the stellar emission phase, but the sources are now the dust grains and peel-off is not performed. The absorbed luminosities are stored in another table, the absorbed dust luminosity table. 
This procedure is repeated in a series of cycles, until convergence of the thermal emission field has been achieved.

\noindent
\paragraph{Dust emission phase} Now that the amount of absorbed luminosity in the dust system has converged, the dust emission step calculates the emission spectra one last time and subsequently launches dust photon packages to determine the final emergent radiation field, including the emission by dust particles. During this last step, peel-off is enabled to determine the contribution of the dust emission to the measured flux.

\noindent
\paragraph{Writing} During the last phase, among other things, the result of the simulation is written out in the form of text files for the spectral energy distributions (SEDs) or FITS files for the data cubes (simulated multi-wavelength images).

\section{SKIRT parallelization modes}
\label{modes.sec}
\vspace{0.5em}


\subsection{Use cases}
\vspace{0.5em}

\noindent
Besides a certain flexibility for accommodating the wide variety in computing systems (as shown in section \ref{parallelization.sec}), we also want SKIRT to accommodate various use cases or applications. A first important distinction is that between \emph{oligochromatic} and \emph{panchromatic} simulations. Oligochromatic simulations are used to calculate the effect of scattering and absorption by dust on the emission of the sources in one or a few specific wavelengths (UV and optical). Due to the limited sampling of the radiation field in wavelength space, re-emission by the dust is not included. Panchromatic simulations, on the other hand, operate on a range of wavelengths that spans the entire relevant spectrum from UV up to submillimetre (submm). These simulations are therefore used to analyse the emission by dust particles. Panchromatic simulations can yield datacubes and SEDs over the entire wavelength spectrum.

\vspace{0.5em}
To optimize the design of the hybrid parallelization, we divided the use cases of SKIRT into four different categories.

\paragraph{Basic oligochromatic simulations} Some applications require only oligochromatic simulations. In general, these simulations have a limited memory usage and a limited runtime. A specific application where (a large number of) low-resolution oligochromatic simulations are performed is within FitSKIRT.

\paragraph{Advanced oligochromatic simulations} When only the transmission of radiation in UV and optical wavelengths is required, but the resolution in wavelength space or physical space is high. An example could be a study of the UV attenuation in a star forming region. Other oligochromatic models are more demanding because they require a high number of photon packages, for example simulations of high optical depth models such as the slab model as part of the TRUST benchmark (Gordon et al., submitted).

\paragraph{Modest-resolution panchromatic simulations}
All applications where galaxies or dusty objects are studied in the infrared, require panchromatic simulations. These simulations need a fair number of wavelength points to resolve the photon-dust interaction, but they do not necessarily require a high-resolution spatial grid and/or a large number of photon packages. In some cases, this is because the underlying models are coarse \citep[e.g.][]{CampsFIRpropEAGLE2016}, and in other cases because only SEDs have to be calculated \citep[e.g.][]{2016MNRAS.458.2288S}. Simulations such as these usually fit on a single node and are often part of a batch where a set of quantities has to be calculated for a large number of models.

\paragraph{High-resolution panchromatic simulations}
Some panchromatic simulations require such a high spatial and spectral resolution that they don't fit in the memory of a single node, and/or require such a large number of photon packages that the runtime on a single node is unacceptably long. Examples include the simulation of high-resolution SPH input models such as ERIS \citep{Saftly2015} or detailed 3D analyses of the dust heating mechanisms in existing galaxies \citep[e.g.][]{VIAENE2016}.

 


\subsection{Parallelization modes}
\vspace{0.5em}

\noindent
To accommodate the various applications and computer architectures, we have conceptualized three different parallelization modes for SKIRT: \emph{multi-threaded}, \emph{hybrid task parallel} and \emph{hybrid} \emph{task+} \emph{data} \emph{parallel} mode. The mode that involves no parallelization is called \emph{single-threaded} mode. The different modes listed below more or less correspond to the different use cases listed above, in the same order.

\noindent
\paragraph{Single-threaded mode} In single-threaded mode, the simulation is executed \emph{serially}, using only one logical core at a time.

\noindent
\paragraph{Multi-threaded mode} In multi-threaded or single processing mode, only one process is launched (no MPI), so all work is divided amongst memory-sharing threads. Obviously, this setup requires a shared-memory system (such as a single node). Although the precise number varies from one system to another, pure multithreading in SKIRT scales well only up to 8-12 cores. Beyond this, speedups are marginal and whatever additional gain doesn't justify the extra occupation of resources. By adding more and more threads, at some point the simulation will actually be slowed down because all the threads simultaneously try to access the same data. For small jobs, even a limited number of parallel threads can provide a sufficiently short runtime. Another use case can be simulations run on a personal computer (where the number of cores is limited anyway and installing MPI would not be very meaningful) or running on systems where an MPI installation is not available.



\noindent
\paragraph{Hybrid task parallel mode} In hybrid task parallel mode, the multithreading capabilities are supplemented with the distributed-memory component (MPI). This mode is useful for simulations too computationally demanding to be run on a personal computer or on a single node of a dedicated computing system. It provides excellent load balancing by assigning each thread within each process to a large number of \emph{chunks}; essentially batches of photon packages of a particular wavelength. Each process (and each thread) performs only part of the calculation for the emission, but does so across the entire wavelength range. Because of this design, the task parallel mode is also highly suited for simulations with fewer wavelengths (e.g. oligochromatic simulations). A downside of this method is that the simulated model is replicated on each process, which drives up the memory consumption on the computing nodes.


\noindent
\paragraph{Hybrid task+data parallel mode} To solve issues with memory consumption for high-resolution simulations, the hybrid task+ data parallelization was implemented as an extension of the `plain' hybrid task parallelization. In this mode, the computational domain of the simulation is split along the wavelength dimension and along the spatial (dust cell) domain. This is then combined with the task assignment of the task parallel mode, with the important difference that processes are always assigned to a subset of the wavelengths. For reasons of load balancing, this parallelization mode is only suitable for simulations with a large number of wavelengths. For these cases, the memory consumption per process is drastically reduced compared to the task parallel mode, while providing the same performance and scaling (in some cases even better due to more efficient communications). The task+data parallelization mode is also useful for high-resolution simulations that do not fit on a node (in multi-threaded or hybrid task mode), even when the number of wavelengths is not particularly high, although this may come at the cost of suboptimal load balancing.

\section{Design and implementation}
\label{design_and_implementation.sec}

\subsection{Design}
\vspace{0.5em}

\begin{figure}[t!]
\centering
\includegraphics[width=0.4\textwidth]{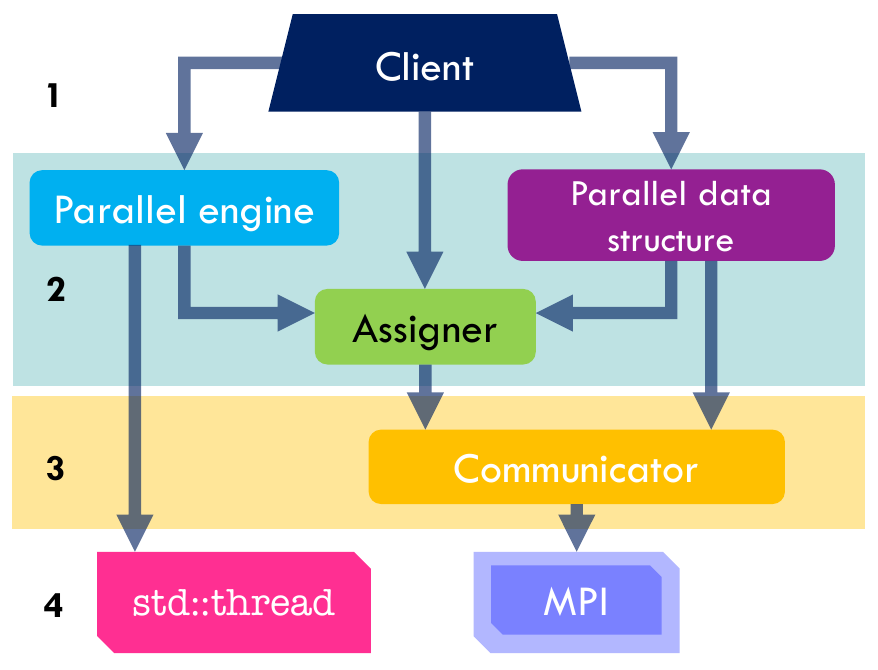}
\caption{The abstraction of the parallelization as implemented in SKIRT. The layer that is visible to the client is composed of the parallel engine object, the assigner and the parallel data structure. The communicator represents a separate layer between the assigner and parallel data structure on the one hand and the MPI library on the other hand. The parallel engine depends on the assigner, as well as on the C++ threading library.}
\label{abstraction.fig}
\end{figure}

\noindent
We present the abstraction scheme that describes the design of the hybrid task+data parallelization in SKIRT. 
The abstraction consists of four separate layers, as visualized in Figure \ref{abstraction.fig}. While designed in the context of a MCRT code, other codes or algorithms may benefit from the adoption of a similar scheme. We list the different components from highest to lowest abstraction level. 

\subsubsection{The client}
\vspace{0.5em}

\noindent
The \emph{client} (or the \emph{caller}) can be any part in the code where parallelization may be useful. In the case of a MCRT code, the client can for example be the photon-shooting algorithm. The desired behaviour (executing a bunch of tasks in parallel to enhance the performance) can be achieved by a small number of calls to the underlying abstraction layer (the interface).

\subsubsection{The interface}
\vspace{0.5em}

\noindent
The interface to the client is provided by three components: the \emph{parallel engine}, the \emph{assigner(s)} and the \emph{parallel data structure(s)}. The parallel engine represents the core of the task-based parallelization. The data-parallel component of the abstraction is represented by the \emph{parallel data structures}. The assigner serves as a base to both the parallel engine and the parallel data structures and thus provides the link between the task-based and data parallel aspects. 
The assigner lays out the plan of which parallel process is assigned to which part of the work. The division of the workload into tasks, as well as the initialization of the assigner, is the responsibility of the client. 

\vspace{0.5em}
The parallel engine uses the assigner to schedule the tasks to the appropriate execution cores until all calculations have been performed. One call to the parallel engine from the client is sufficient to invoke the parallel execution of a predefined set of tasks.

\vspace{0.5em}
A parallel data structure is an object that represents a data structure central to the algorithm and thus not tied to one individual process. In other words, it is a shared resource in which the different parallel entities can modify and access information. Since multiprocessing is a form of distributed memory programming, strictly speaking data is not shared between the processes. Keeping the data synchronized between the processes requires communications (either over the interconnections between sockets or over the node network). The goal of the parallel data structure is to provide an interface for such a data structure and to keep it synchronized: the parallel data structure is responsible for all communications that are necessary to achieve this. The synchronization happens under the hood so that for the client, the use of the parallel data structure is no different from the use of a resource within a shared-memory context. The parallel data structure thus essentially mimics the behaviour of genuinely shared data structures.

\vspace{0.5em}
Encapsulating the relevant information in parallel data structures has another advantage: the parallel data structure can anticipate which part of the data is required for which process at any given time, and change its internal representation accordingly. More specifically, the object can take advantage of the distributed memory layout and let processes store only a subset of the data, that which is required for the tasks it is assigned to at that time. When new tasks are assigned that require different parts of the data, the internal representation changes  (the data is reshuffled). 

\subsubsection{The communicator}
\vspace{0.5em}

\noindent
The \emph{communicator} is the object that represents the multiprocessing environment. 
The communicator is used by the assigners to determine between how many processes the tasks can be distributed. The communicator is also solely responsible for handling all of the communications between processes, so the parallel data structure uses it to keep its data set synchronized.

\subsubsection{The external dependencies}

\vspace{0.5em}
\noindent
The base of the abstraction scheme is represented by the external software libraries that are used for the multithreading and multiprocessing capabilities. These libraries are supported on a wide variety of  systems. 

 
\subsection{Implementation}
\label{implementation.sec}
\vspace{0.5em}

\noindent
Each of the different constituents of the abstraction scheme described above, not counting the external libraries, can be identified with either a class or a set of similar classes in the SKIRT code. The dependencies between these (groups of) classes have been indicated in Figure \ref{abstraction.fig} by means of downwards pointing arrows. The core element in the abstraction, the parallel engine, is represented in SKIRT by the \mytexttt{Parallel} class, which provides a `single-method' interface to the rest of the SKIRT code (the client). For its multithreading capabilities it directly uses the standard \mytexttt{std::thread} class. For the multiprocessing functionality it depends on the \mytexttt{ProcessAssigner} and \mytexttt{ProcessCommunicator} classes (and their respective subclasses) in the first place. In the bottom layer, the MPI library has an additional encapsulation in SKIRT, provided by the \mytexttt{ProcessManager} class. In our discussion of the implementation of each layer, we will move up through the abstraction graph, starting with the \mytexttt{ProcessManager} class.


\subsubsection{The \texttt{ProcessManager} class}
\label{manager.sssec}
\vspace{0.5em}

\noindent
The \mytexttt{ProcessManager} class, indicated in figure \ref{abstraction.fig} as a layer around MPI, is intended to provide an suitable interface to the C MPI library. This class mainly exists to ensure that there are no direct MPI calls spread over the code. Instead, all MPI calls must pass through the \mytexttt{ProcessManager} class, which provides an alternative single-process implementation when MPI is not available on the system. The \mytexttt{ProcessManager} class contains only static methods, hence no instances of \texttt{ProcessManager} can be created. 

\vspace{0.5em}
In a SKIRT simulation, all processes cooperate in the MCRT algorithm. Thus, within the SKIRT runtime context, there exists only one multiprocessing environment (or \emph{process pool}) which can be only used once at any given time. This means that only one client ought to acquire the process pool for its purposes. Therefore, we have implemented the \mytexttt{ProcessManager} class so that it hands out the MPI resource such that this particular condition is always met. An \emph{atomic} integer is stored by this class to count the number of active requests for MPI. The requests are handled by the \mytexttt{acquireMPI} and \mytexttt{releaseMPI} methods that are in the public interface of \mytexttt{ProcessManager}. The \mytexttt{acquireMPI} method increments the request counter and checks whether the MPI environment is still available. If this is the case, the \mytexttt{ProcessManager} sets some variables for the caller so that the latter knows its own process rank and the number of processes in the environment. The \mytexttt{releaseMPI} method decrements the requests counter, so that the multiprocessing environment can be passed on to the next caller. Clients calling \mytexttt{acquireMPI} in the meantime are not granted access to the process pool.


\vspace{0.5em}
In SKIRT, MPI is requested during the setup of a particular radiative transfer simulation and subsequently released when that simulation finishes. The pattern described above is useful when multiple of these simulations are run in parallel, e.g. when using the automatic fitting routine FitSKIRT. In this case, it is the fitting scheme that requests MPI first, so that individual simulations are always executed within one process even when a multiprocessing environment is available.

\subsubsection{The process communicators}
\label{communicators.sssec}
\vspace{0.5em}

\begin{figure}[t!]
\centering
\includegraphics[width=0.40\textwidth]{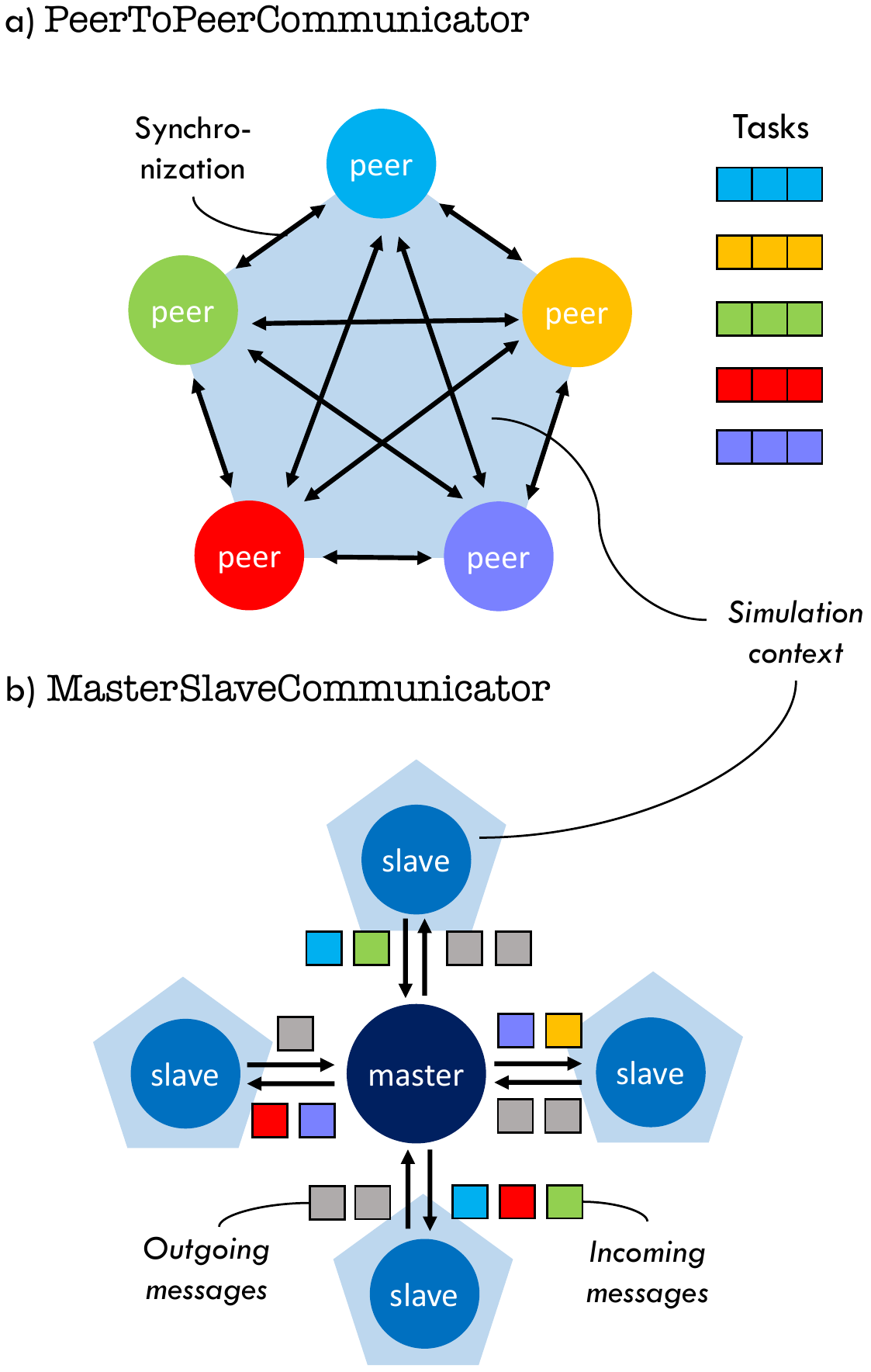}
\caption{The two kinds of communicator models. Top: the peer-to-peer model; Bottom: the master-slave model. The peers in the peer-to-peer model are part of the same simulation, in which they are each assigned to specific tasks. Communication between all the peers is required to keep the data structures synchronized. In the master-slave model, a slave receives a message from the master, performs some work according to the contents of that message, and returns the relevant results. The data structures created by the different slaves are completely independent: one slave does not need to know the state of the other slave's memory.}
\label{communicators.fig}
\end{figure}

\noindent
A process communicator is used to represent an ensemble of processes and to implement the communications that can be performed between them. In SKIRT, two different kinds of communicators have been implemented, designed to serve very different purposes. The first kind, the \mytexttt{PeerToPeerCommunicator} is used for the parallelization of a single SKIRT simulation. It defines a whole range of communications that are decentralized, involving the participation of each process. During a SKIRT simulation, these communication methods are used in order to make sure that all the essential data structures are properly synchronized between the different processes. Hence, the \texttt{PeerToPeerCommunicator} class is optimized for a parallelization scheme where the general flow or progress of the program is inherently dependent on the entire program state, which must be shared by all parallel processes (or at least for a significant part). This is illustrated in Figure \ref{communicators.fig}. To facilitate this synchronization process, the assignment of tasks within the simulation to the different parallel processes is \emph{predetermined} (by the assigners).

\vspace{0.5em}
The \mytexttt{MasterSlaveCommunicator}, on the other hand, is designed for parallelizing the execution of many independent simulations, each with their own data structure. One process is assigned the \emph{master}, whose responsibility is delegating the other processes (which are called the \emph{slaves}). Communications implemented by the \mytexttt{MasterSlaveCommunicator} are always directed towards or away from the master. The \mytexttt{MasterSlaveCommunicator} is perfectly suited for the FitSKIRT routine, where a large number of similar models are simulated in a parallelized loop. Since these simulations are completely independent, there is no need for synchronizing data structures across the parallel processes and communications between the slaves are therefore nonexistent. Instead, the slaves dynamically get assigned new tasks by the master, by means of small messages. In the case of FitSKIRT, these messages describe the parameters required for setting up a radiative transfer simulation. After a process finishes executing such a simulation, it sends a message back to the master process with the relevant results.

\subsubsection{The process assigners}
\label{assigners.sssec}
\vspace{0.5em}

\begin{figure}[t!]
\centering
\includegraphics[width=0.3\textwidth]{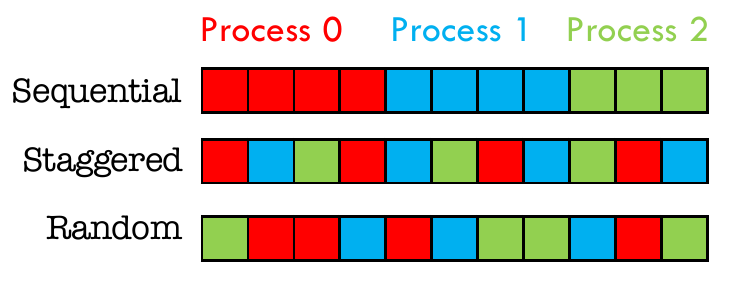}
\caption{An illustration of how the different process assigners distribute a hypothetical set of 11 tasks amongst 3 parallel processes (colors represent different processes).}
\label{assigners.fig}
\end{figure}

\noindent
As explained before, the peer-to-peer multiprocessing model relies on a prior knowledge of which process executes which part of the work. Determining which process is assigned to which part is the task of a process assigner. \mytexttt{ProcessAssigner} is the base class that defines a common interface for the different assigner classes in SKIRT. Figure \ref{assigners.fig} illustrates how the different assigners distribute a given set of tasks amongst the process pool. The different \mytexttt{ProcessAssigner} subclasses are: 

\paragraph{SequentialAssigner} The \mytexttt{SequentialAssigner} divides the work into sequential sections and then assigns each process to one such section. 
A \mytexttt{SequentialAssigner} can be used to parallelize the execution of a set of tasks which show no systematic variation in load.


\paragraph{StaggeredAssigner} The \mytexttt{StaggeredAssigner} distributes the work in a \emph{round-robin} or cyclic fashion: the first task is given to the first process, the second task to the second process and so on, until the pattern is repeated. The number of tasks assigned to any process by the \mytexttt{StaggeredAssigner} is identical to the number of tasks assigned thereto by the \mytexttt{SequentialAssigner}. 
The \mytexttt{StaggeredAssigner} provides a significantly better load balancing than the \mytexttt{SequentialAssigner} in those cases where there is a systematic variation (gradient) in the load of the tasks. In SKIRT, the execution time for the life cycle of a photon packages depends on the wavelength, because photons in the UV and optical regime are far more likely to scatter compared to infrared and submm photons. Also, adjacent dust cells can be expected to have a similarly strong radiation field, leading to a spatial correlation in the computation time of the dust emission spectra. This is the reason that the \mytexttt{StaggeredAssigner} has been adopted as the default task assigner in SKIRT for these purposes.



\paragraph{RandomAssigner} The \mytexttt{RandomAssigner} uses random numbers to decide which process is assigned to a certain task. For each individual task, a random number is drawn uniformly from zero to the number of processes in the communicator. The resulting assignment scheme is synchronized between the different processes. The purpose of the \mytexttt{RandomAssigner} is, like \mytexttt{StaggeredAssigner}, to obtain good load balancing when the tasks have unequal workload. Although providing an equally good load balancing compared to the \mytexttt{StaggeredAssigner} during testing (at least for a large number of tasks), the latter has been used as the default assigner in SKIRT.




\subsubsection{The \texttt{Parallel} class}
\label{parallel.ssec}
\vspace{0.5em}

\begin{figure}[t!]
\centering
\includegraphics[width=0.45\textwidth]{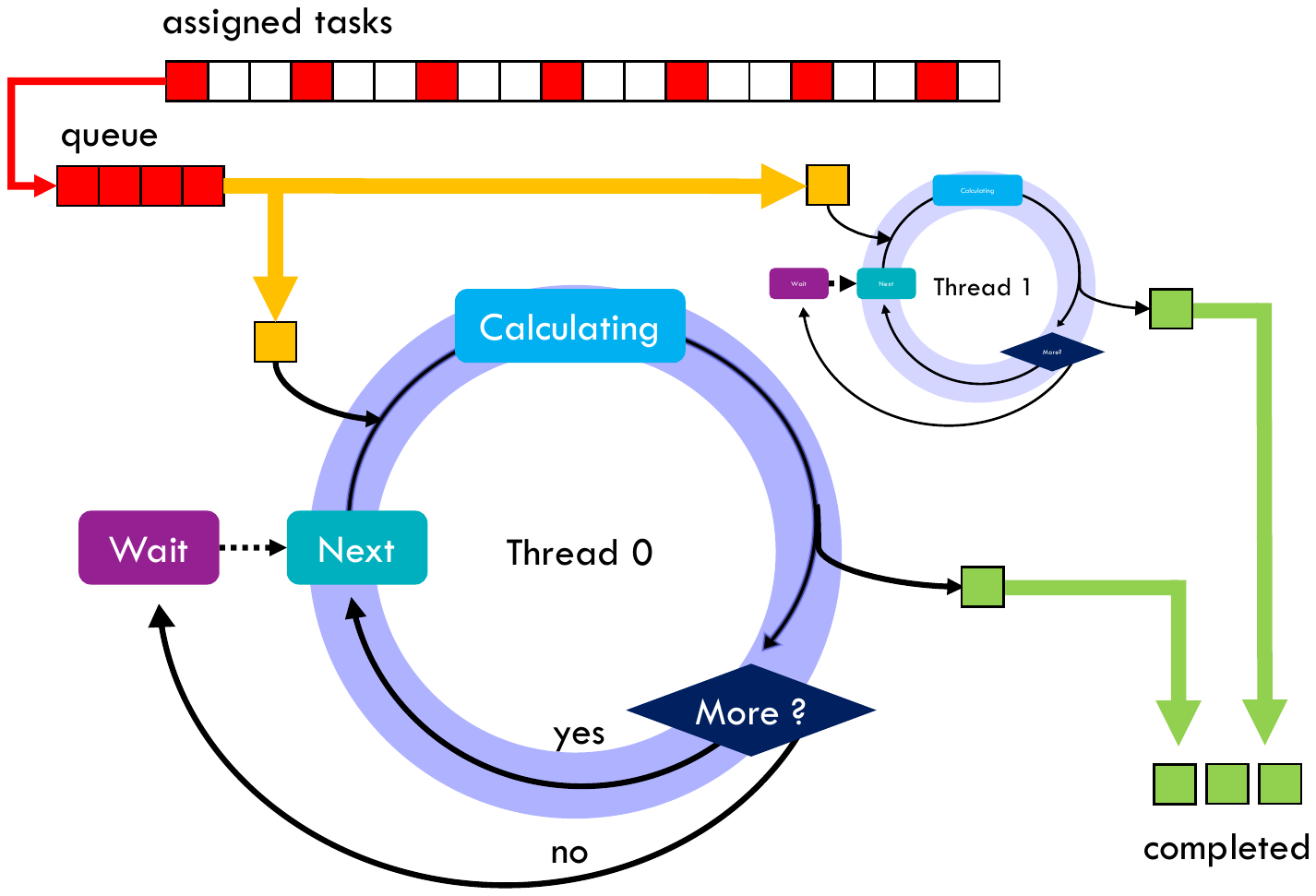}
\caption{The lifecycle of a thread in SKIRT. In this example, a process contains two execution threads (thread 0 on the foreground and thread 1 in the background). Tasks assigned to the process are stacked into a queue, where they await execution by one of two threads. Which thread executes which specific task is determined by random events. When a thread is ready, a task is picked from the queue and execution of the task is initiated. For as long as there are tasks left, the thread remains in a loop where repeatedly the next task in the queue is performed. When no more tasks are left in the queue, the thread is put in a wait condition, until it is recycled for a new set of tasks.}
\label{threadlife.fig}
\end{figure}

\noindent
The core mechanisms of the hybrid task parallelization in SKIRT are implemented in the \mytexttt{Parallel} class. It can be identified with the parallel engine in the abstraction of Figure \ref{abstraction.fig}. This class has no dependencies to the main SKIRT code beyond the components in the abstraction. Its external dependencies include the standard \mytexttt{std::thread} class and related classes required for locking and wait conditions. The multiple execution threads are created when the \mytexttt{Parallel} object is constructed and can be reused from one parallel section of the code to another. The \mytexttt{Parallel} object can therefore be said to represent the pool of available threads, while it is also aware of the multiprocessing environment. 

\begin{figure}[t!]
\centering
\includegraphics[width=0.45\textwidth]{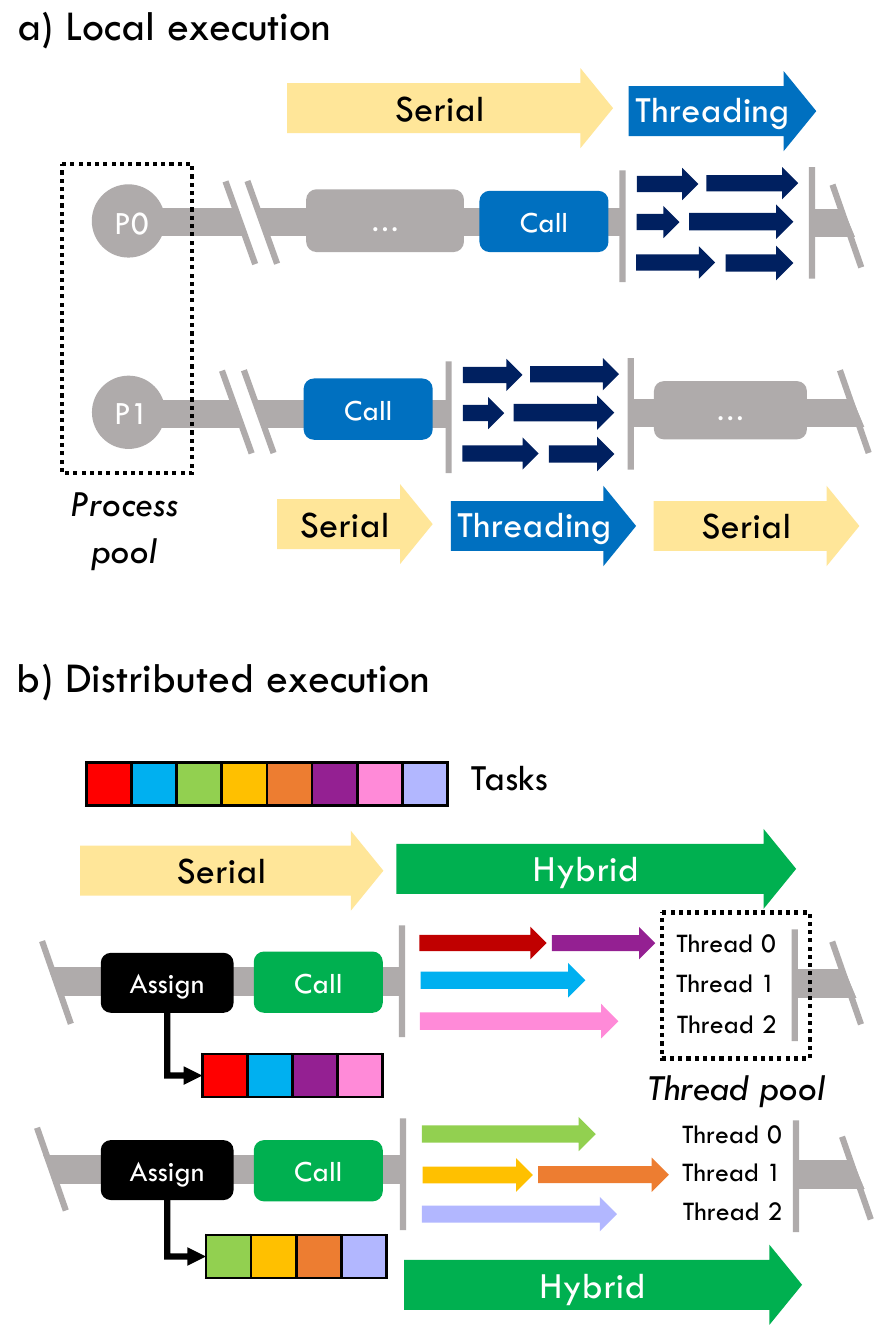}
\caption{Schematic depiction of the hybrid of multi-threaded execution of a set of tasks in SKIRT, in the case of two processes and three threads per process. Top panel (a): at some point in the algorithm, process P1 requires the execution of a certain number of tasks (six in this example). The output is neither influenced by nor relevant for process P0. Process P1 therefore executes the \mytexttt{call} function in local mode, which invokes the multi-threaded execution of the set of tasks. At another point in time, process P0 also uses its thread pool in a similar way to speed up its computations. Bottom panel (b): a set of eight tasks, tied to the global state of the simulation, has to be performed. To speedup the calculation, the tasks are distributed amongst the different processes using an assigner. Each process invokes the \mytexttt{call} function providing this assigner to utilize its thread pool for the execution of the assigned tasks. The result is the hybrid parallelization of the original set of tasks amongst the computing units.}
\label{hybrid.fig}
\end{figure}

\vspace{0.5em}
Figure \ref{threadlife.fig} illustrates the lifecycle of a thread within a \mytexttt{Parallel} object. Upon creation, the thread is placed under a wait condition. When work for the thread arrives, the thread is woken up and the parallel execution is started. At that point, the thread enters a loop in which a specific \emph{target} is called repeatedly. The target can be a function, or an object that contains a \mytexttt{body} method. The target body takes the task index and executes the appropriate portion of work. An index is incremented \emph{atomically} each time a thread starts a new iteration. Without further locking, this atomicity can guarantee that each task is executed exactly once. Across processes, this condition is naturally guaranteed by the assignment scheme. This is illustrated in Figure \ref{hybrid.fig} (b) for two processes, each using three threads. The total workload consists of 8 distinct tasks. When the assigner is created, it fixes the mapping from these tasks to the different processes. The parallel execution of the tasks is then initiated when the method \mytexttt{call} of the \mytexttt{Parallel} instance is invoked on each process. Threads will concurrently execute the tasks assigned to their parent process, taking the next task - if available - when they finish the previous one. While the tasks assigned to a process are fixed a priori, the work is distributed between the threads within that process dynamically, providing the best possible load balancing between threads.

\vspace{0.5em}
The \mytexttt{call} function can be invoked in two different ways. As a first argument, it always takes the target. If the call function is provided with an assigner as the second argument, this invokes the hybrid parallelization (\emph{distributed} execution) for the target function. Threads will only get handed out tasks that were assigned to the parent process. For \emph{local} execution (multi-threaded execution of work tied to just a single process), the \mytexttt{call} function also accepts the length of the task queue and subsequently hands out these tasks amongst the threads. This is illustrated in Figure \ref{hybrid.fig} (a). 

\vspace{0.5em}
An example from the SKIRT code of the use of the \mytexttt{Parallel} class for hybrid parallelization is presented below, to show the effectiveness of the abstraction. 


{
\small
\begin{verbatim}
parallel->call(this, &MonteCarloSimulation::
               dostellaremissionchunk,
               _lambdagrid->assigner(), _Nchunks);
\end{verbatim}
}

\vspace{0.5em}
In this case, the body function that is supposed to be called in parallel is the \mytexttt{dostellaremissionchunk} function of the \mytexttt{MonteCarloSimulation} class, which shoots photon packages for one chunk in a particular wavelength. The number of tasks to be executed is the number of wavelengths, held by the wavelength assigner of the simulation (\mytexttt{\_lambdagrid->assigner()}). The assigner is passed as the third argument to the \mytexttt{call} function, and the last argument of the function determines how many times each task should be repeated (the number of chunks per wavelength \mytexttt{\_Nchunks} in this case).





\subsubsection{The \texttt{ParallelTable} class}
\label{paralleltable.ssec}
\vspace{0.5em}

\noindent
The first of two parallel data structures implemented in SKIRT is the \mytexttt{ParallelTable} class. An instance of this class represents a two-dimensional container (elements are accessed by a row and a column index) of double-precision values. The data structure can be represented as $N_{\text{c}}$ columns and $N_{\text{r}}$ rows. As already indicated in Figure \ref{skirtflow.fig}, SKIRT contains three such data structures: two tables of absorption luminosities and one table of emission spectra. In all cases, the dimension of the table is \emph{number of wavelengths} $\times$ \emph{number of cells}. 


\vspace{0.5em}
The \mytexttt{ParallelTable} has a few key features. First of all, it makes an abstraction of the underlying data structures so that element access is consistently provided through a row index $i$ (the dust cell index) and a column index $j$ (the wavelength index), in that order. Under the hood, the data can be stored in different representations, depending on the situation. This is the second feature of the \mytexttt{ParallelTable}: its internal behaviour depends on the parallelization mode.

\paragraph{Multithreading mode} In multithreading (single processing) mode, the ParallelTable will allocate a sequence of $N_{\text{c}} \times N_{\text{r}}$ double-precision values. The pair $(i,j)$ is internally converted to element $i\times N_{\text{c}} + j$ of this sequence. The state of the table is naturally shared amongst all threads, but threads must write data atomically to prevent race conditions. Synchronization or communication is not necessary.

\paragraph{Hybrid task parallel mode} In hybrid task parallel mode, all processes (and threads) simulate all wavelengths and thus write data to each column of the absorption table. Similarly, during dust emission phases, dust emission luminosities ought to be available for each wavelength and for each dust cell. This requires the absorption tables to be synchronized after each emission cycle and the emission spectra tables to be synchronized after each emission spectra calculation step. In hybrid task parallel mode, the \mytexttt{ParallelTable} organizes itself in such a way that each process contains the entire sequence of $N_c \times N_{\text{r}}$ values (as in multithreading mode), but it also keeps corresponding elements of the table on different processes synchronized at the appropriate times. Because the absorbed luminosity in a dust cell adds linearly, this is achieved by summing the tables of the different processes element-wise when a cycle completes.

\begin{figure}[tb]
\centering
\includegraphics[width=0.45\textwidth]{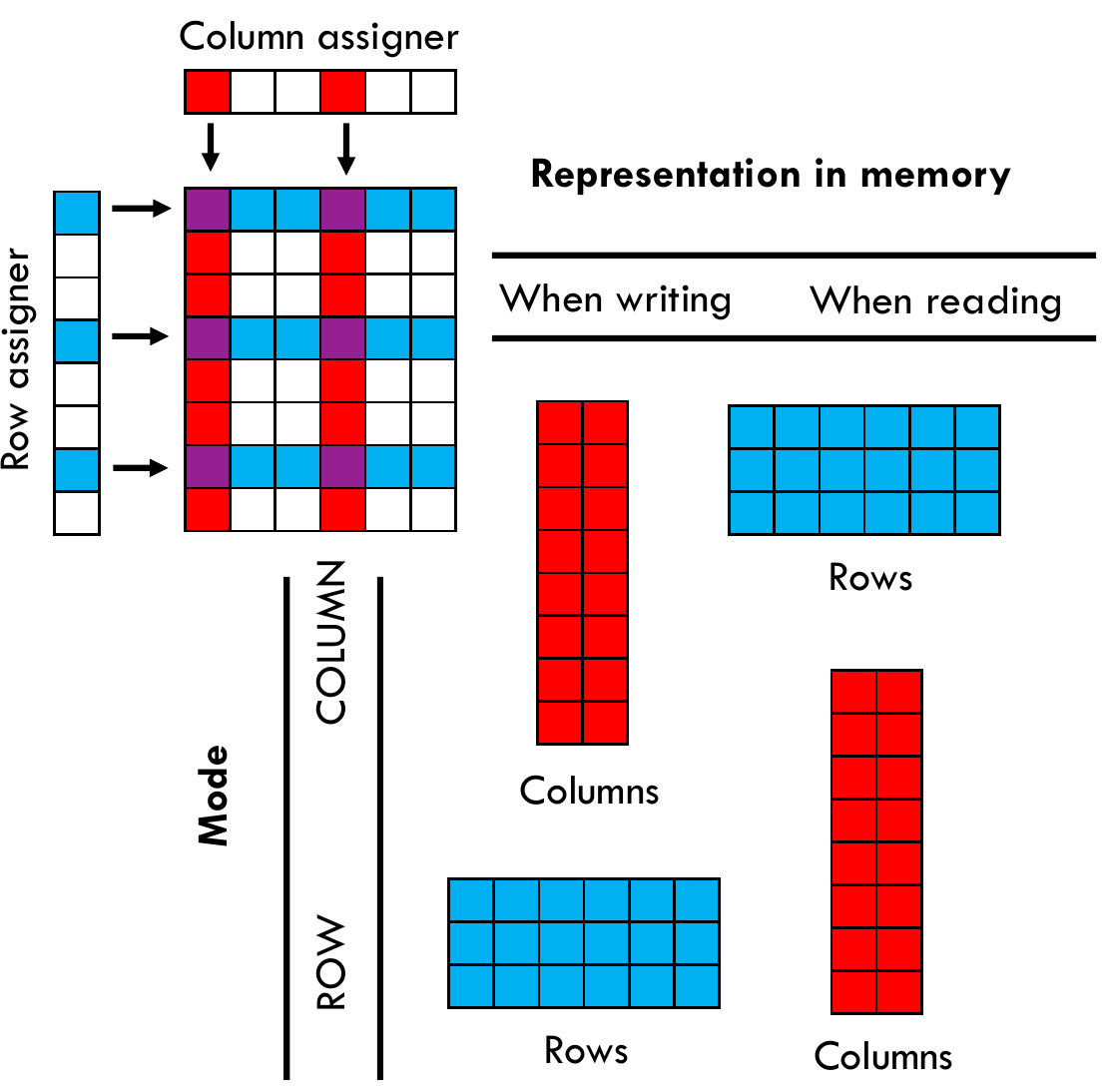}
\caption{The initialization of a \mytexttt{ParallelTable} object from a column and row assigner. Depending on whether the table is setup in \mytexttt{ROW} or \mytexttt{COLUMN} mode, the internal representation of the table is different. A new parallel table in \mytexttt{COLUMN} mode is represented as a subset of its columns, as dictated by the column assigner. Data can be written to the table if it belongs in one of these columns. When elements have to be read out from the table, the table has to be changed in representation. The table is now represented by a subset of the rows, as prescribed by the row assigner. The case of a parallel table initialized in \mytexttt{ROW} mode is analogous.}
\label{paralleltable.fig}
\end{figure}

\begin{figure*}[!htb]
\centering
\includegraphics[width=0.95\textwidth]{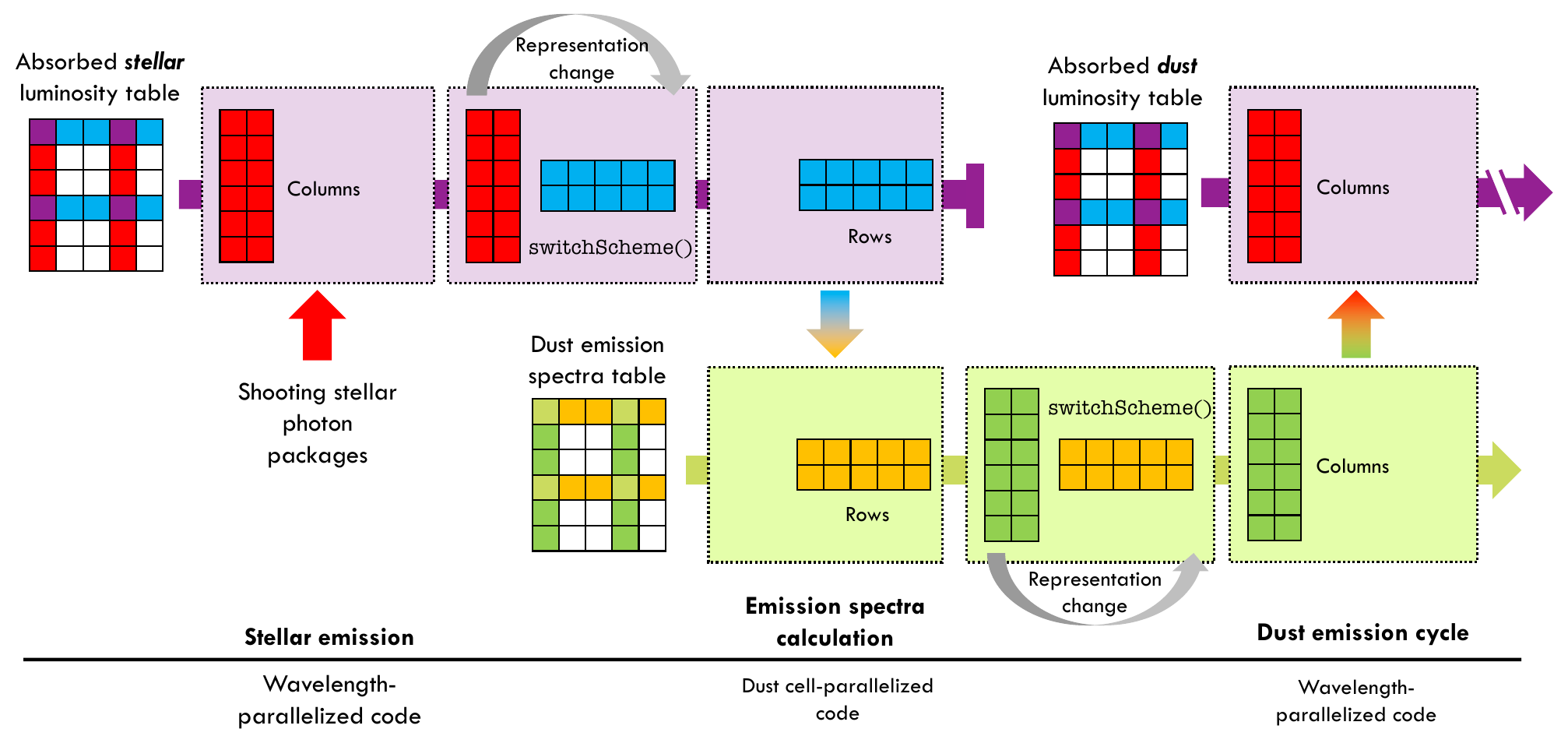}
\caption{The representation changes of the parallel tables in SKIRT, in task+data parallel mode. When photon packages are launched during the stellar emission phase, the table of stellar absorption luminosities is stored in column representation. This means that luminosities can be stored for all dust cells but only for a subset of wavelengths. On each process, this subset and thus the specific column representation is different. After the stellar emission phase, the table is converted to row representation. Once again, the representation on each process is different since different processes are assigned different dust cells. The \mytexttt{switchScheme} method of the parallel table arranges all necessary communications to reshuffle the data across processes. After the representation switch, the table can be read from. During the next phase, dust emission spectra are calculated based on the absorption spectra stored in the rows of the absorbed stellar luminosity table. The resulting spectra are stored in the dust emission spectra table, also a parallel table. This table, however, is in row representation during the time it is written to, and changes representation when information must be read from its columns. These columns, containing the emission luminosities of the entire dust grid at a specific wavelength, are used during the dust emission cycles, where the shooting of dust photon packages is parallelized over the wavelengths.}
\label{representations.fig}
\end{figure*}

\paragraph{Hybrid task+data parallel mode} When all wavelengths are simulated by all processes, task parallellization cannot be complemented by a data parallelization because partitioning the tables along any axis would induce a need for continual communications to bring the right portion of data to the right process. When data parallelization is enabled, however, a wavelength assigner of the type \mytexttt{StaggeredAssigner} is created. This wavelength assigner and the dust cell assigner are passed to the constructor of the \mytexttt{ParallelTable} class as the column and row assigner respectively. This is illustrated in Figure \ref{paralleltable.fig}. Also specified in the constructor is whether \mytexttt{ROW} or \mytexttt{COLUMN} representation should be used for writing data to the table. The absorption table, constructed in \mytexttt{COLUMN} mode, will then only allocate a subset of the columns (wavelengths). The emission spectra tables, constructed in \mytexttt{ROW} mode, will allocate a subset of the rows (dust cells). This is switched under a representation change. In the case of the absorption table, the column representation is used because photons are launched only for specific wavelengths at a given process. The emission spectra table uses the row representation because the calculation of emission spectra is parallelized over dust cell indices. The same assigners that are used for the tables are evidently used by the \mytexttt{Parallel} class to schedule the tasks and thus provide the framework for the task+data parallel mode to work.

\vspace{0.5em}
Because of the division in columns or rows, when using $N_{\text{p}}$ processes, the memory requirement for the absorption or emission table is reduced to $1/N_{\text{p}}$ of that in plain task parallel mode. Yet, to the client, these tables can still be treated as a regular 2D (non-distributed) table: all index conversions are handled internally by \mytexttt{ParallelTable}. 

\vspace{0.5em}
An additional feature of the \mytexttt{ParallelTable} in task+data parallel mode is that it can switch representations. This representation change is the equivalent of the synchronization of the table in task parallel mode. After a (stellar or dust) emission cycle, emission spectra have to be calculated per individual dust cell, thus this requires that complete rows in the absorption table can be accessed. Similarly, dust emission cycles require emission luminosities for all dust cells (but only some wavelengths), so complete columns of the emission table are required. In other words, the absorption tables need a shift from \emph{column} representation to \emph{row} representation and the emission table needs a shift from \emph{row} representation to \emph{column} representation. This transposition is implemented internally by \mytexttt{ParallelTable} by means of a collective communication operation that shuffles information around so that each process obtains the required data.

\vspace{0.5em}
Figure \ref{representations.fig} demonstrates the representation changes that the parallel tables in SKIRT undergo during a panchromatic simulation. The representation change is invoked by calling the \mytexttt{switchScheme} method of \mytexttt{ParallelTable}, after an emission phase in the case of the absorption table and after the emission spectra calculation for the emission spectra table.

\vspace{0.5em}
Due to an efficient implementation of the communication, the row and column representation must only exist simultaneously for a short interval on any process. Because this does not need to occur at the same time for all three tables, the memory scaling per process due to the data parallelization is essentially $1/N_{\text{p}}$. For large simulations, the size of the parallel tables dominate the total memory usage of SKIRT, whereby the total memory requirement of the simulation can be expected to stay at a constant value in the limit of many processes. Thus, both the runtime and the memory usage per process decrease when processes are added. For comparison, in task-based mode the total memory requirement increases more or less linearly with the number of processes.



\subsubsection{The \texttt{ParallelDatacube} class}
\label{paralleldatacube.ssec}

\vspace{0.5em}
\noindent
The other parallel data structure in SKIRT is the \mytexttt{ParallelDatacube}. An instance of this class represents a data cube (one wavelength axis and two spatial axes) of which the wavelength slices can be distributed across the different processes. Similarly as \mytexttt{ParallelTable}, it behaves as a 2D table, but in this case indexed on wavelength and pixel number. Its internal structure also depends on the parallelization mode: when in multithreading or task parallel mode, slices are not distributed but the entire cube is stored (at each process) and synchronization (if necessary) is achieved by collective summation. In hybrid task+data parallel mode, the wavelength assigner is used to determine its reduced representation and index conversions and for assembling the complete datacube at the end of the simulation (using a collective gather operation). The memory scaling is the same as for the \mytexttt{ParallelTable} ($1/N_{\text{p}}$). In both hybrid modes, the synchronization or assembly is performed by calling the \mytexttt{constructCompleteCube} method. Similar as for the parallel tables, when one datacube has to be assembled (to be written out), the other parallel datacubes can retain their minimal representation, limiting the strain on the memory as much as possible.


\subsubsection{Lockfree additions}
\label{lockfreeimplementation.ssec}
\vspace{0.5em}


\begin{figure}[tb]
\centering
\includegraphics[width=0.43\textwidth]{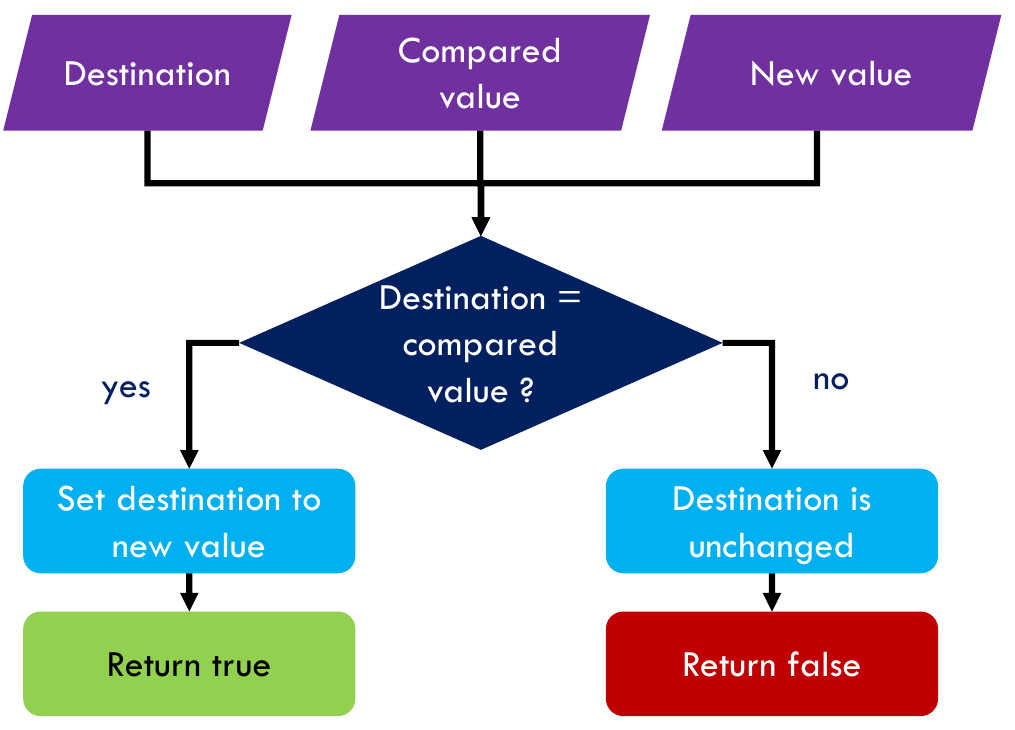}
\caption{A flowchart illustrating the mechanism of the compare-and-swap (CAS) instruction. Three variables are involved: the destination variable, the value to compare the destination with, and the variable that contains the new value for the destination. CAS checks whether the destination value matches the compared value. If this is the case, the destination value is replaced by the new value. In the other case, the destination is left unchanged. To notify the caller whether the operation has succeeded, the CAS function returns true (1) or false (0) respectively.}
\label{cas.fig}
\end{figure}

\begin{figure*}[!ht]
\centering
\includegraphics[width=0.9\textwidth]{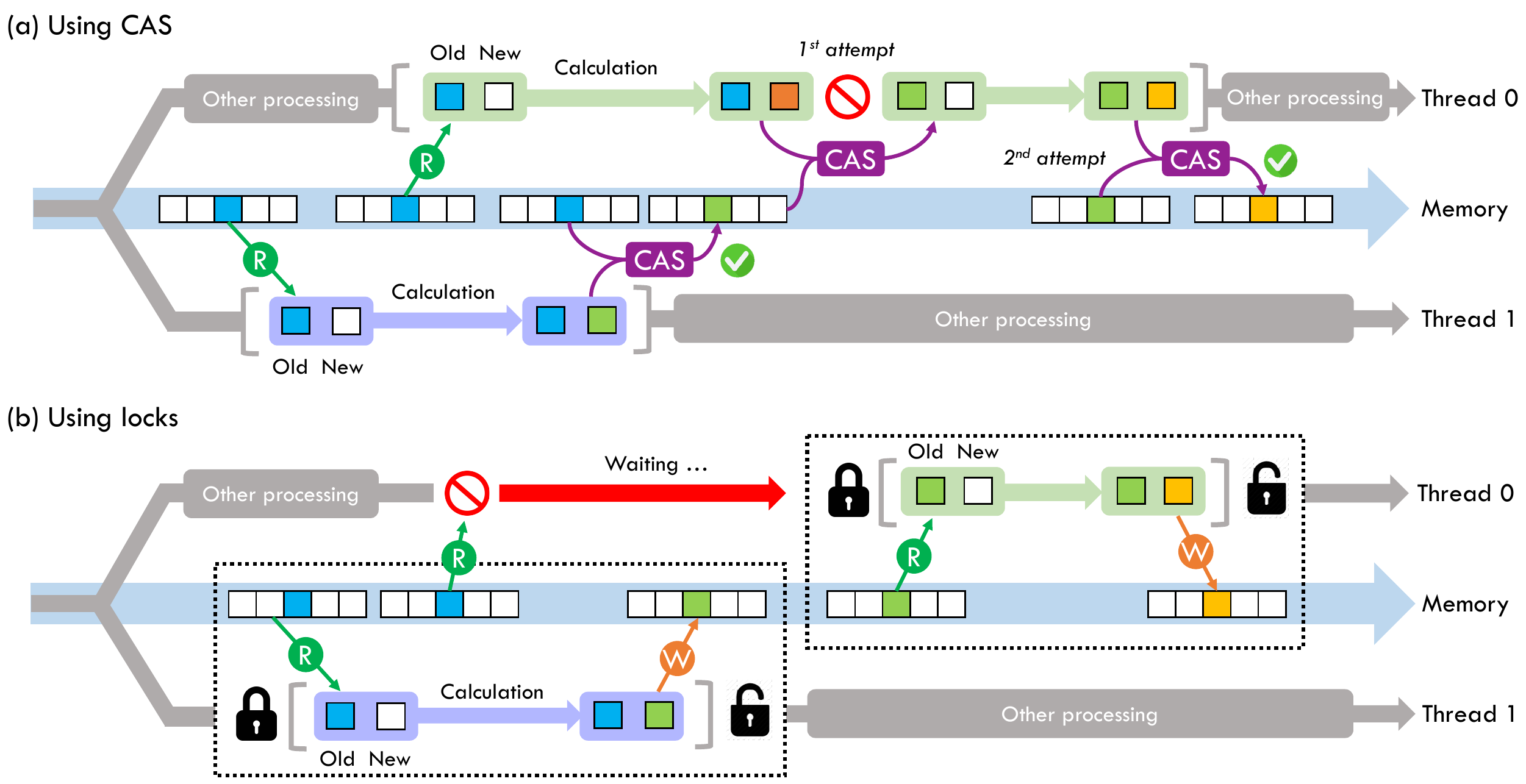}
\caption{Contention for a shared variable between two parallel threads. Top panel (a): writing to the shared variable is implemented with a call to the compare-and-swap (CAS) function. The \mytexttt{LockFree::add} function embeds the CAS call inside a loop, which performs multiple attempts to overwrite the variable if necessary. The atomicity is ensured by the CAS function, the variable is only locked at the finest instruction level. Bottom panel (b): the contention is solved by using locks. The first thread to arrive, acquires the lock and can safely change the shared variable while the other thread is halted. When the variable is changed by thread 1 and the lock released, the lock is acquired by the other process.}
\label{lockfree.fig}
\end{figure*}

\noindent
In SKIRT, the absorption tables are constantly updated as luminosities are added to a certain dust cell and wavelength bin. As discussed in section \ref{lockfreeprogramming.ssec}, there are various problems associated with using high-level locks to synchronize these additions between multiple threads. First of all, acquiring a lock takes time. This time depends on the specific implementation and on the operating system, and may require a kernel call (i.e. require the action of the operating system) even if the lock is not contended. In any case, it requires multiple instructions. Secondly, a lock also uses some memory, which is important to consider when implementing them. In SKIRT, using an individual lock for each element in the data structures would increase the memory requirements unacceptably. When, on the other hand, one single lock would be used for an entire table, there would be so much contention that every thread would be suspended for most of its CPU cycles. Therefore, we would be forced to implement complex schemes that limit the number of locks (and thus the number of acquisitions) but at the same time suppress contention by keeping the locks relatively fine-grained. 

\vspace{0.5em}
Lock-free operations, on the contrary, allow solving this problem with a minimal set of very fast instructions. On most current platforms, there is no hardware-supported atomic double addition (in contrast to atomic integers). Instead, we use the atomic \emph{compare-and-swap} operation (CAS) \citep{Herlihy1991} operation. The compare-and-swap operation can be utilized for any write operation to shared data in concurrent environments, not just for a simple addition as is the case in SKIRT.

\vspace{0.5em}
Since C++11, a CAS function is available from the standard \mytexttt{std::atomic} library. The atomic CAS instruction is supported in hardware by current multiprocessor architectures. Non-atomic variants of the CAS operation are used in lock-based contexts. The mechanism of CAS is depicted in Figure \ref{cas.fig}. CAS is a conditional write operation: a memory location is only overwritten by a new value if its current value matches a `control' value. Most often, this control value or comparison value is the value of the destination variable read in by the thread before calling the CAS operation. A thread that has to update the memory location retrieves the current value, and based on this calculates the new value. The old value and the new value are passed to the CAS function, which checks whether the old value still represents the current state. If it doesn't, another thread was able to change the memory contents in the meantime. Checking the values and subsequently changing the memory contents happens atomically, so that it is guaranteed that the memory location is not altered by another threads between these two steps. 

\vspace{0.5em}
Thus, the CAS operation can either fail or succeed. In the case of SKIRT, CAS fails for a thread that has calculated the increased value of the luminosity in a dust grid cell during a peel-off event or a detection event in an instrument pixel, if another thread has succesfully increased this value as well in the meantime. For the first thread, the calculation of the new value has to be redone. When contention is high, yet another thread may have altered the luminosity within this short timespan. In this case, subsequent calls to the CAS function are necessary until it is succesfull, for example in a loop.

\vspace{0.5em}
To facilitate the use of consecutive CAS calls in a loop, we have created a \mytexttt{LockFree} namespace in which we have implemented an \mytexttt{add} function. The upper panel of Figure \ref{lockfree.fig} illustrates how the \mytexttt{Lockfree::add} function is used in SKIRT. In this case, two concurrent threads both try to change the state of a certain memory location within the same timeframe. Thread 1 does so slightly before thread 0, which results in the first thread having calculated a new (updated) value first as well. With the variable still unchanged in the shared data sequence, thread 1 succesfully performs the compare-and-swap operation, replaces the old value (blue) by the new value (green). By that time, thread 0 has finished its calculation and attempts the CAS operation to change the state of the variable (to orange). This attempt fails, however, because its recollection of the original value doesn't match the present state anymore. The 'old' value held by thread 0 is then replaced by the present value, and this thread calculates a new updated value. The second attempt to change the memory state succeeds in this example since no other thread did so in then meantime. At the end, while a CAS operation might not always succeed from the first attempt, both threads have changed the state of the shared variable in a proper manner.

\vspace{0.5em}
In the bottom panel of Figure \ref{lockfree.fig}, an alternative situation is pictured where locking is used instead. The initial condition is identical: both threads arrive at a point where they need to update a certain shared variable but the one thread arrives slightly before the other. The resource where both threads are now competing for is the lock: the first one to acquire it stalls all other threads at the point of acquisition. Thread 1 can now safely change the state of the variable because any thread that would attempt the same will be suspended. After the new state is written, the critical section is left and the lock is released. At that point, thread 0 is acknowledged in its attempt at acquiring the lock and performs the same steps as thread 1.




\vspace{0.5em}
The code snippet below shows how the absorption event is implemented in SKIRT using the \mytexttt{LockFree::add} function. A photon package in the wavelength bin \mytexttt{ell} increases the absorbed luminosity in the dust cell with index \mytexttt{m} with a value of \mytexttt{DeltaL}. This value is added atomically to the element \mytexttt{(m, ell)} of the 2D parallel table \mytexttt{Labsvv}.

{
\small
\begin{verbatim}
void DustSystem::absorb(int m, int ell, double DeltaL)
{
    LockFree::add(Labsvv(m,ell), DeltaL);
}
\end{verbatim}
}

\vspace{0.5em}
The implementation of the \mytexttt{add} function (in the \mytexttt{LockFree} namespace) beats locking implementations not only in performance but also in simplicity:

{
\small
\begin{verbatim}
inline void add(double& target, double value)
{
    auto atom = new(&target) std::atomic<double>;
    double old = *atom;
    while(!atom->compare_exchange_weak(old,old+value))
    {}
}
\end{verbatim}
}

\vspace{0.5em}
The \mytexttt{LockFree::add} statement is all that is necessary to ensure thread safety for an ordinary addition statement. Adopting this lock-free mechanism has shown to improve the performance of the photon shooting algorithm in SKIRT significantly. To our knowledge, lock-free mechanisms are not commonly adopted yet in scientific codes.

\section{Using SKIRT with parallelization}
\label{skirtwithparallelization.sec}
\vspace{0.5em}

\subsection{Command-line}
\label{commandline.ssec}
\vspace{0.5em}


\noindent
The parallelization mode and hybridization (processes, threads) can be specified on the command line. Whether hybrid mode can be used depends of course on whether MPI is installed. If it is, SKIRT can be launched through the \mytexttt{mpirun} (or \mytexttt{mpiexec}) tool to enable multiprocessing. The number of processes is specified by the \mytexttt{-n} option of \mytexttt{mpirun}. 
The SKIRT executable takes the other arguments that are relevant for the parallelization. The option \mytexttt{-t} option is used to specify the number of threads. Note that in hybrid mode (multiprocessing enabled), this is the number of threads \emph{per process} (so the actual number of cores that should be allocated is $N_{\text{p}} \times N_{\text{t}}$). The number of threads used by SKIRT defaults to the number of logical cores on the system, which is why it is important to specify a more appropriate number of threads in the hybrid parallelization modes. To enable the hybrid task+data parallelization, the \mytexttt{-d} flag must be added. Thus, the command for launching a SKIRT simulation in hybrid task+data parallel mode would look like this:


\begin{verbatim}
mpirun -n <np> skirt -t <nt> -d <skipath>
\end{verbatim}

\vspace{0.5em}
A complete list of the command-line options that can be specified for SKIRT is listed on the SKIRT website \footnote{\hyperlink{www.skirt.ugent.be}{www.skirt.ugent.be}}.

\vspace{0.5em}
Note that \mytexttt{mpirun} also accepts a range of other command-line options, that can be relevant for SKIRT. The task parallelization in SKIRT works best if a process from the pool is assigned cores that share `local' access to memory, i.e. on the same die or socket. Processes should not be spread across sockets or nodes but instead be \emph{bound} to dies, sockets, or (if necessary) nodes. Furthermore, the assignment scheme should be identical for each process to ensure proper load balancing (each process must have the same amount of resources). There are a handful of options for the \mytexttt{mpirun} command that can produce such setups. Users are referred to the \mytexttt{mpirun} documentation for a complete list \footnote{See \hyperlink{https://www.open-mpi.org/doc/v2.0/man1/mpirun.1.php}{https://www.open-mpi.org/doc/v2.0/man1/mpirun.1.php} for a list of mpirun options for the OpenMPI implementation}. Other options can be available depending on the implementation and the version of MPI. Different options may also be available through the job scheduling system, if applicable.

\subsection{The Python front-end}
\vspace{0.5em}

\noindent
In an effort to make it more convenient for the SKIRT user to exploit the hybrid parallelization on multi-core machines, we have developed a front-end as part of our PTS (Python Toolkit for SKIRT) package \footnote{\hyperlink{http://www.skirt.ugent.be/pts/}{http://www.skirt.ugent.be/pts/}}. With the proper configuration, users can launch SKIRT simulations on any remote machine using a one-line command. Another command gives an overview of the status of scheduled simulations, whether they are still running, finished or have not yet started. The results of finished simulations are automatically transferred back to the client computer and the remote copy of the output folder is removed. The only prerequisite on the remote side is a working SKIRT executable and support for the SSH and SFTP protocols.

\vspace{0.5em}
As part of this PTS front-end, we implemented a tool that generates the most optimal hybrid parallelization scheme for a particular SKIRT simulation. When using the PTS simulation launcher, the optimal parallelization scheme is automatically determined and used for the job. The algorithm uses the properties of the system's architecture and the current load on the system to optimize the simulation for the best performance.


\section{Tests and results}
\label{results.sec}
\vspace{0.5em}

\subsection{Hardware}
\vspace{0.5em}


\noindent
SKIRT has been tested thoroughly on various systems, with and without the hybrid parallelization. For the test results presented in this work, however, we used the same infrastructure to perform the simulations. This way, it is easier to compare results and draw conclusions. We used the Tier-2 infrastructure of Ghent University (Stevin) \footnote{\hyperlink{http://www.ugent.be/hpc/en}{http://www.ugent.be/hpc/en}}, partner of the Flemish Supercomputer Centre (VSC) \footnote{\hyperlink{https://www.vscentrum.be/}{https://www.vscentrum.be/}}, with a total capacity of 174 TFlops (8768 cores over 440 nodes). It consists of 5 different clusters. We used the \mytexttt{Delcatty} and \mytexttt{Swalot} clusters for running the tests because they are best equipped for multi-node jobs. They also are the largest parallel machines available to us, and offer a reproducible performance. Using their scheduling system, we could allocate resources (nodes) without experiencing interference from other users. The \mytexttt{Delcatty} cluster consists of 160 nodes with two 8-core Intel E5-2670 CPU packages, and with 64 GB of memory. The \mytexttt{Swalot} cluster incorporates two 10-core Intel E5-2660v3 packages on each of its 128 nodes. Both clusters have an FDR InfiniBand network to connect their nodes.

\subsection{Test model}
\vspace{0.5em}

\noindent
For our performance tests, we used a panchromatic simulation of a spiral galaxy with a flattened Sersic bulge of stars in the center, a ring of stars and dust, and an exponential disk of stars and dust with two spiral arms (produced by an analytical perturbation). The wavelength grid ranges from 0.1 to 1000 $\mu m$, and the number of wavelength points was varied from 120 to 160 in the different tests. The dust system was discretized using a Cartesian dust grid with 256 000 dust cells, increased to 900 000 dust cells for the tests that target the communication times. One instrument has been set up that creates an SED and a datacube with 1024 by 1024 pixels (times 160 wavelength planes).


\subsection{Single-node comparison}
\vspace{0.5em}

\begin{figure}[h!]
\centering
\includegraphics[width=0.5\textwidth]{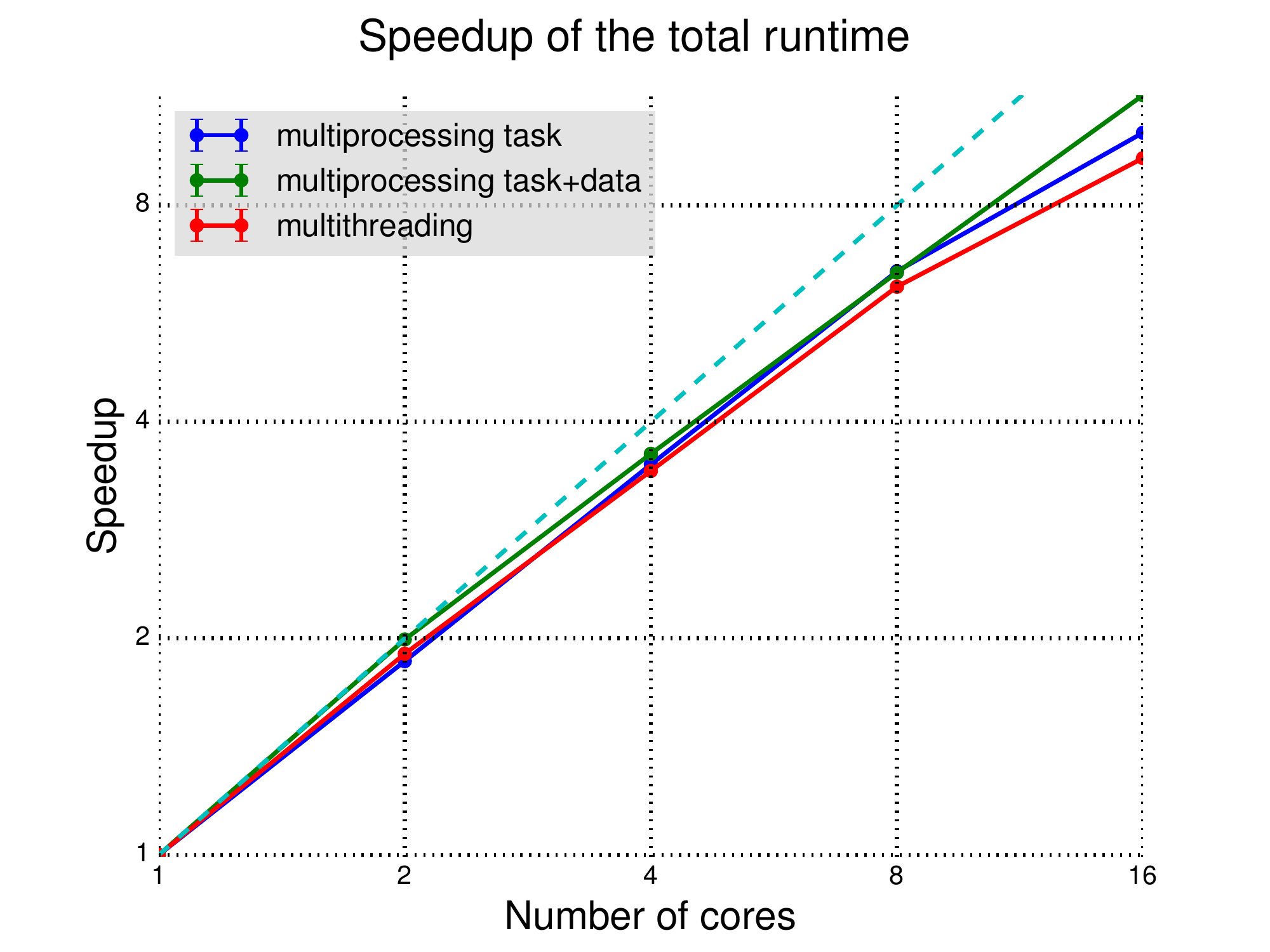}
\caption{The speedup of the simulation as a function of the number of cores on one node (16 cores in total) for three parallelization modes: multithreading, task-based multiprocessing, and task+data multiprocessing. The dashed curve represents the \emph{ideal} speedup, i.e. the speedup of an application that is 100\% parallel without parallelization overhead}
\label{PURE_speedup.fig}
\end{figure}

\noindent
To compare the multithreading and multiprocessing in terms of performance, we conduct a series of tests on a single node. For an increasing number of cores (up to a full node = 16 cores), we record the total runtime of the simulation in pure multithreading mode, and pure multiprocessing mode (with and without data parallelization enabled). The results are shown in Figure \ref{PURE_speedup.fig}. The quantity plotted on the vertical axis is the speedup. The speedup for a certain number of cores $n$ is defined as the ratio between the serial runtime (on one core) divided by the runtime on $n$ cores. Ideal scaling happens if increasing the number of cores speeds up the program by the same factor (a linear increase of the speedup). Comparing the scaling for the three cases, it is clear that pure multithreading is the least efficient on a full node (a speedup of 9 on 16 cores), while the multiprocessing runs reach a speedup of 11 and 10, respectively with and without data parallelization.



\subsection{Load balancing and multi-node scaling}
\vspace{0.5em}

\begin{figure}[h!]
\centering
\includegraphics[width=0.5\textwidth]{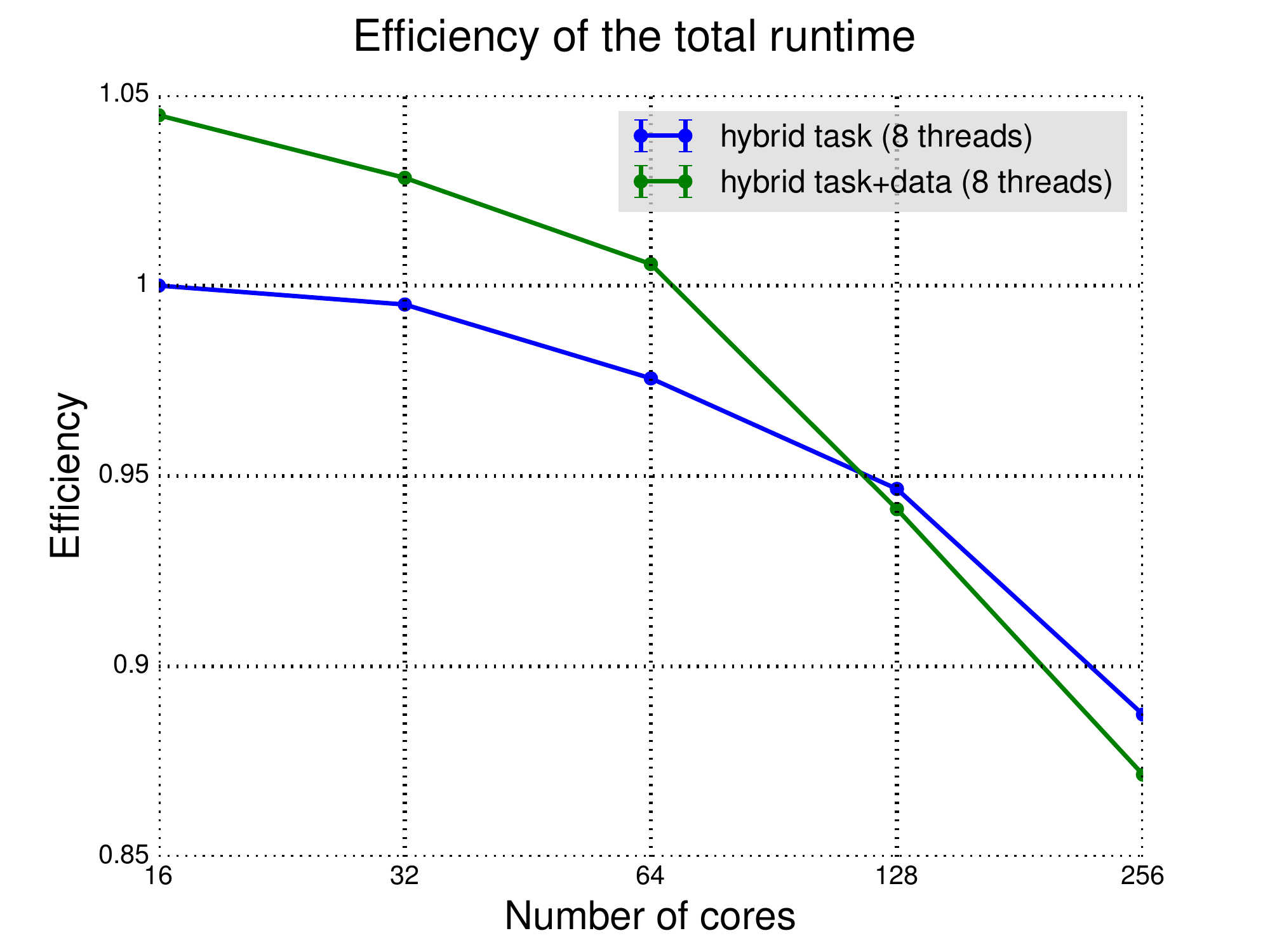}
\caption{The efficiency of the simulation as a function of the number of cores, as measured on multiple nodes. Two processes were used per node, thus the efficiency is plotted from 2 processes up to 32 processes. When comparing the hybrid task+data parallelization and the hybrid task parallelization, it can be seen that the former performs better for a number of processes lower than 16, while for a higher number of processes the task parallel mode gives better performance.}
\label{efficiency_multinodescaling.fig}
\end{figure}

\noindent
To analyse the load balancing, we perform tests with hybrid parallelization on multiple nodes of the cluster. On each node, we assign two processes with eight threads per process and increase the number of nodes from one to 16. Figure \ref{efficiency_multinodescaling.fig} shows the efficiency with and without data parallelization. The effiency is a measure of how efficient the computing resources are used by the parallel algorithm, and is calculated by dividing the speedup by the number of cores.
For this plot, the values have been normalized on the efficiency of the slowest method on one full node (hybrid task parallelization). It is seen that the task+data parallelized simulations perform better up to 4 nodes but beyond that the task parallel mode is faster. This can be explained by considering the load balancing between the parallel processes.


\begin{figure}[h!]
\centering
\includegraphics[width=0.5\textwidth]{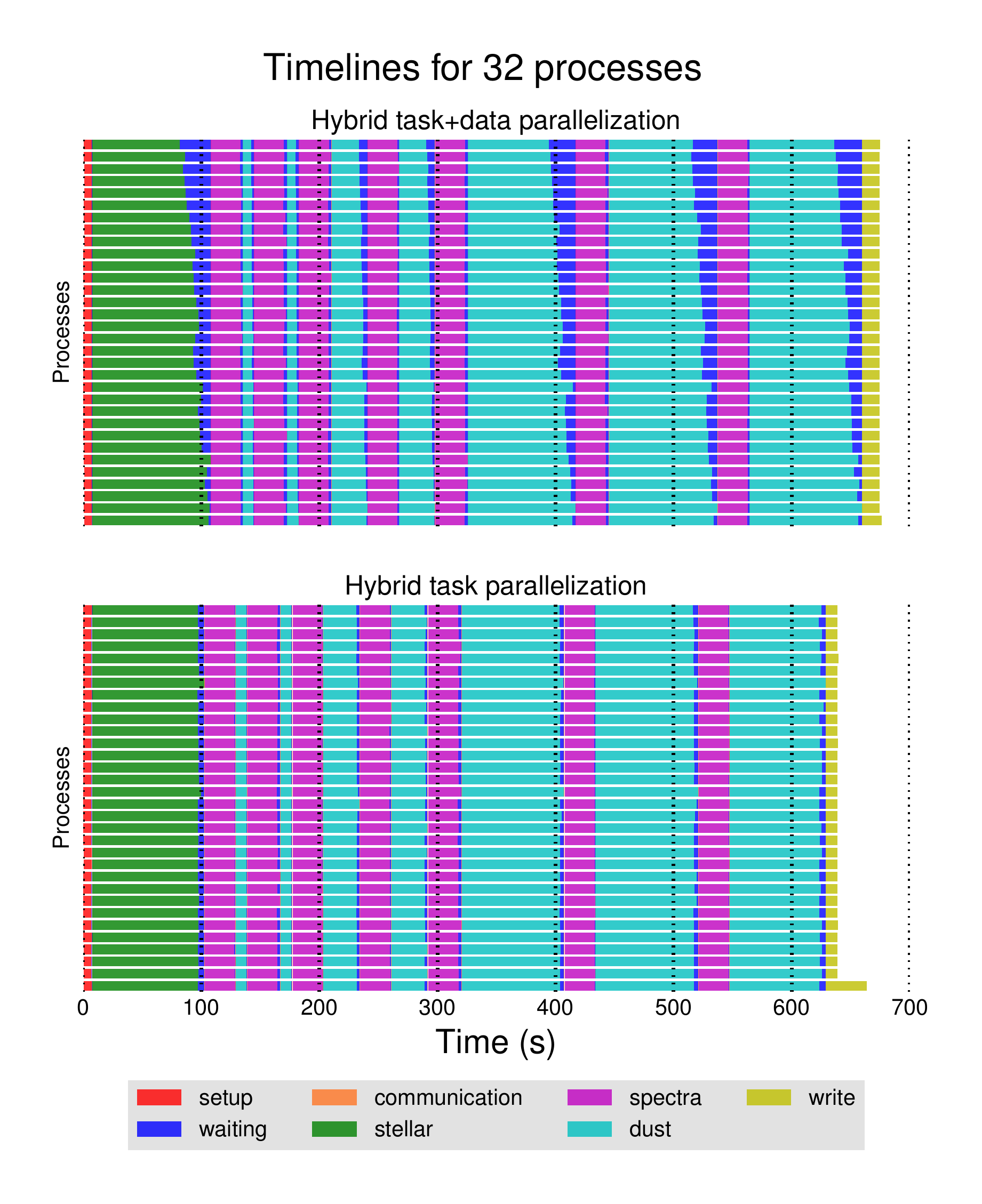}
\caption{Timelines of the different simulation phases. Top: timeline for 32 processes in task+data parallel mode. Bottom: timeline for 32 processes in task parallel mode.}
\label{timeline32.fig}
\end{figure}

\vspace{0.5em}
Figure \ref{timeline32.fig} shows the timelines of the simulations on 16 nodes (corresponding to 256 cores and 32 processes) in task+data parallel mode (top) and `plain' task parallel mode (bottom). Red and yellow bars represent the time spent in serial parts of the simulation (setup and writing). Green, cyan and magenta bars represent the parallel sections: photon package shooting and dust emission spectrum calculations. It is clear from these timelines that the parallelization covers all but a very small part of the simulation runtime. Dark blue bars represent the overhead of the parallelization because of processes that wait for other processes to complete their work (load imbalance). For the task parallel mode, these are insignificant, but they clearly have a negative effect on the performance in task+data parallel mode. Also note that the MPI overhead of communication (plotted in orange) is completely negligible.


\vspace{0.5em}
The load imbalance in task+data parallel mode is a result of distributing the wavelengths amongst the processes for shooting photon packages (green and cyan bars). It is striking that the waiting time increases with process rank. In other words, the processes with the highest ranks finish their tasks (wavelengths) sooner than the processes with lower ranks. This is due to the way wavelengths are assigned by the \mytexttt{StaggeredAssigner}: by iterating over the processes cyclically, the average wavelength assigned to higher-rank processes is slightly higher. Because the number of scattering events per photon decreases with wavelength, the average load is lower for these higher-rank processes. This effect is most prominent with a large number of processes, where the number of wavelengths assigned to each process is small. This result shows that there is an upper limit on the number of processes that can be used for a particular simulation when data parallel mode is enabled. Taking into account the load gradient, other assignment schemes could be devised that improve the load balancing to a small extent. Wavelengths could for example be handed out in pairs, starting with the smallest and largest wavelength.

\subsection{Communication}
\vspace{0.5em}

\begin{figure}[h!]
\centering
\includegraphics[width=0.5\textwidth]{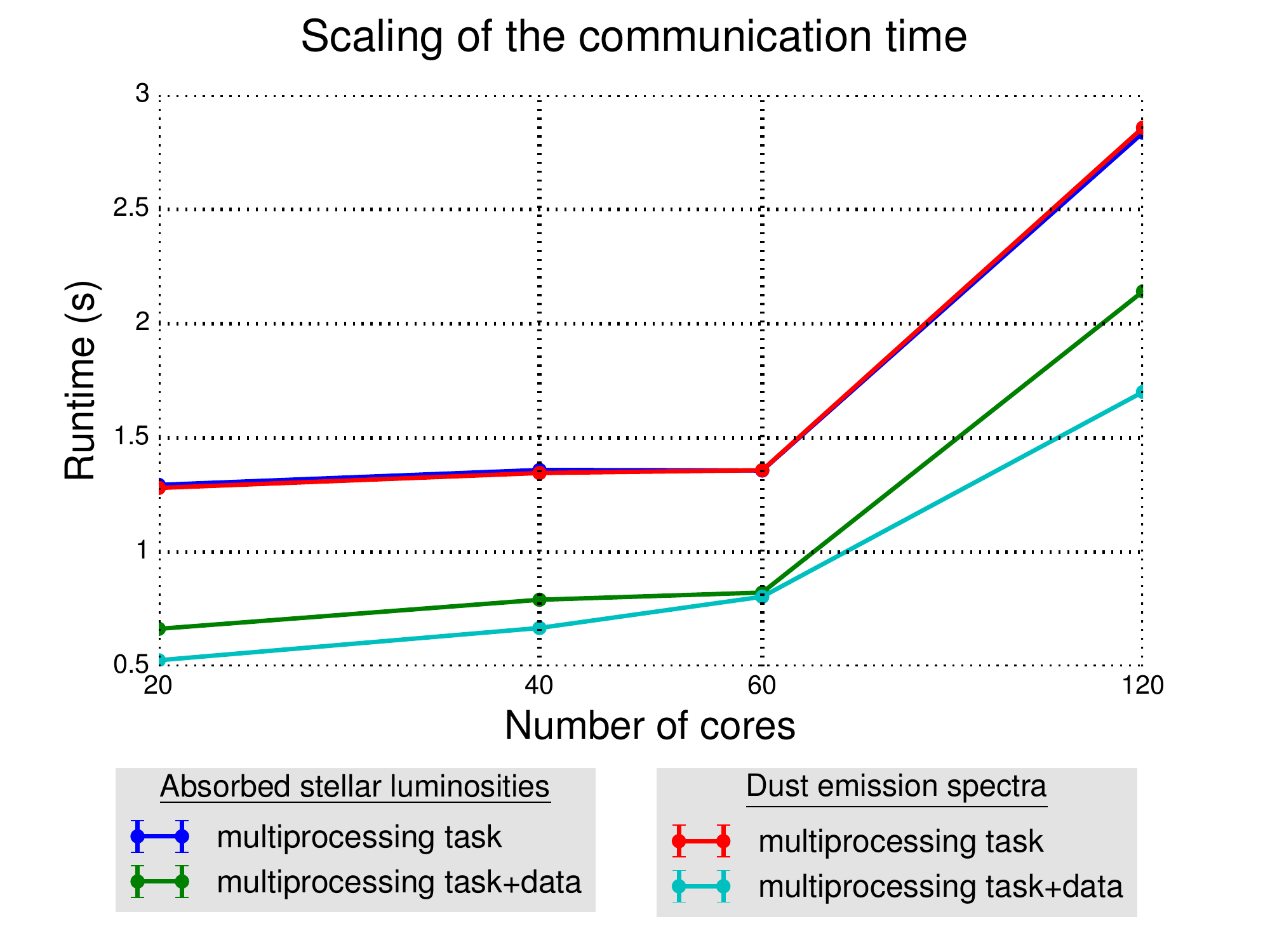}
\caption{The time required for different communication steps in the simulation, as a function of the number of cores (equal to the number of processes in this case). Data points are shown for the synchronization of the absorbed stellar luminosity table and the emission spectra table, in task parallel mode and in task+data parallel mode.}
\label{comparecom.fig}
\end{figure}

\noindent
We set up a batch of simulations on the \mytexttt{Swalot} cluster specifically to time the communication phases for a varying number of processes. We increase the number of single-threaded processes from 20 to 120 (i.e. from one node up to 6 nodes) and record the time required for the synchronization of the absorption and emission table. We do this in both the task parallel mode (where synchronization is achieved through a collective summation across processes) and the task+data parallel mode (where the tables are changed in representation). Figure \ref{comparecom.fig} shows the result of these timings. For any number of processes, the summation of the absorption table and the summation of the emission table take the same amount of time, as expected (upper curves). The transpositions from row to column representation (light blue curve) and from column to row representation (green curve) are also very similar in execution time, but are significantly faster than the summations in task parallel mode. In any case, the communication is not a bottleneck for the performance.


\subsection{Hybridization}
\vspace{0.5em}

\begin{figure}[h!]
\centering
\includegraphics[width=0.5\textwidth]{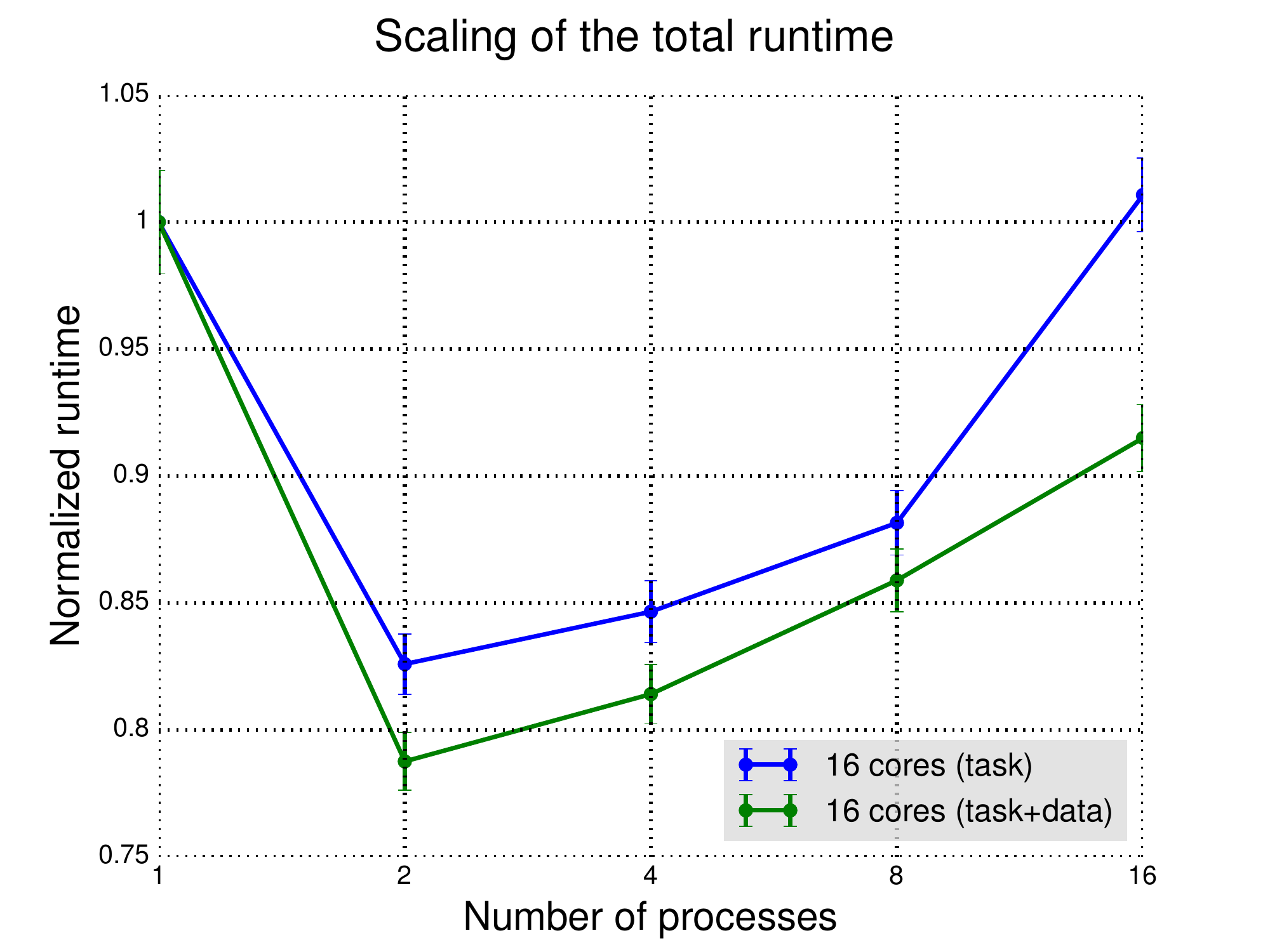}
\caption{Runtime of different runs of the simulation on the same number of cores, but with different combinations of the number of processes and the number of threads per process. The runtimes are normalized to the runtime of the simulation with 1 process and 16 threads.}
\label{hybridefficiency.fig}
\end{figure}

\noindent
The fact that we can choose any hybrid parallelization scheme has multiple benefits. Compared to pure multiprocessing, for example, we can occupy more memory per process and yet still use all available computing cores. Secondly, the overheads of the shared-memory and distributed-memory programming methods can be minimized to obtain the best performance for a given set of resources. This last principle is illustrated in Figure \ref{hybridefficiency.fig}. It shows the normalized runtime of a simulation run on a full node (16 cores), but with different hybridizations. Two curves are shown, one for the hybrid task parallelization and one for the hybrid task+data parallelization mode. In both cases, a scheme of 2 processes per node and 8 threads per process has the shortest runtime. It shows that choosing the best hybrid scheme can improve performance by more than 20\% on this particular system. 


\vspace{0.5em}
The fact that two processes per node seems to be the best strategy can be explained based on the topology of the system. With one process, 16 threads will be sharing the same virtual memory, scattered over the memory of the two sockets. Memory access to other sockets is costly, which explains why the first data point is much higher than the runtimes for two processes, in which case each socket is occupied by a separate process. On the other end of the curve, where each core runs a separate process, the performance depends on the parallelization mode. There is of course a communication overhead, but as shown in Figure \ref{comparecom.fig}, it is negligible (order of a few seconds). The fact that the performance improves when threads are added and the number of processes decreased (towards the middle of the curve), is related to the fact that the load balancing between the processes improves. Because multithreading has proven to scale equally well as multiprocessing up to eight cores on this system (see Figure \ref{PURE_speedup.fig}), the performance keeps improving until both sockets are filled with eight threads. This shows that providing a single `best parallelization strategy' is not possible, but rather depends on the specific hardware in addition to the particular type of SKIRT simulation.

\subsection{Increased number of photon packages}
\vspace{0.5em}

\begin{figure}[h!]
\centering
\includegraphics[width=0.5\textwidth]{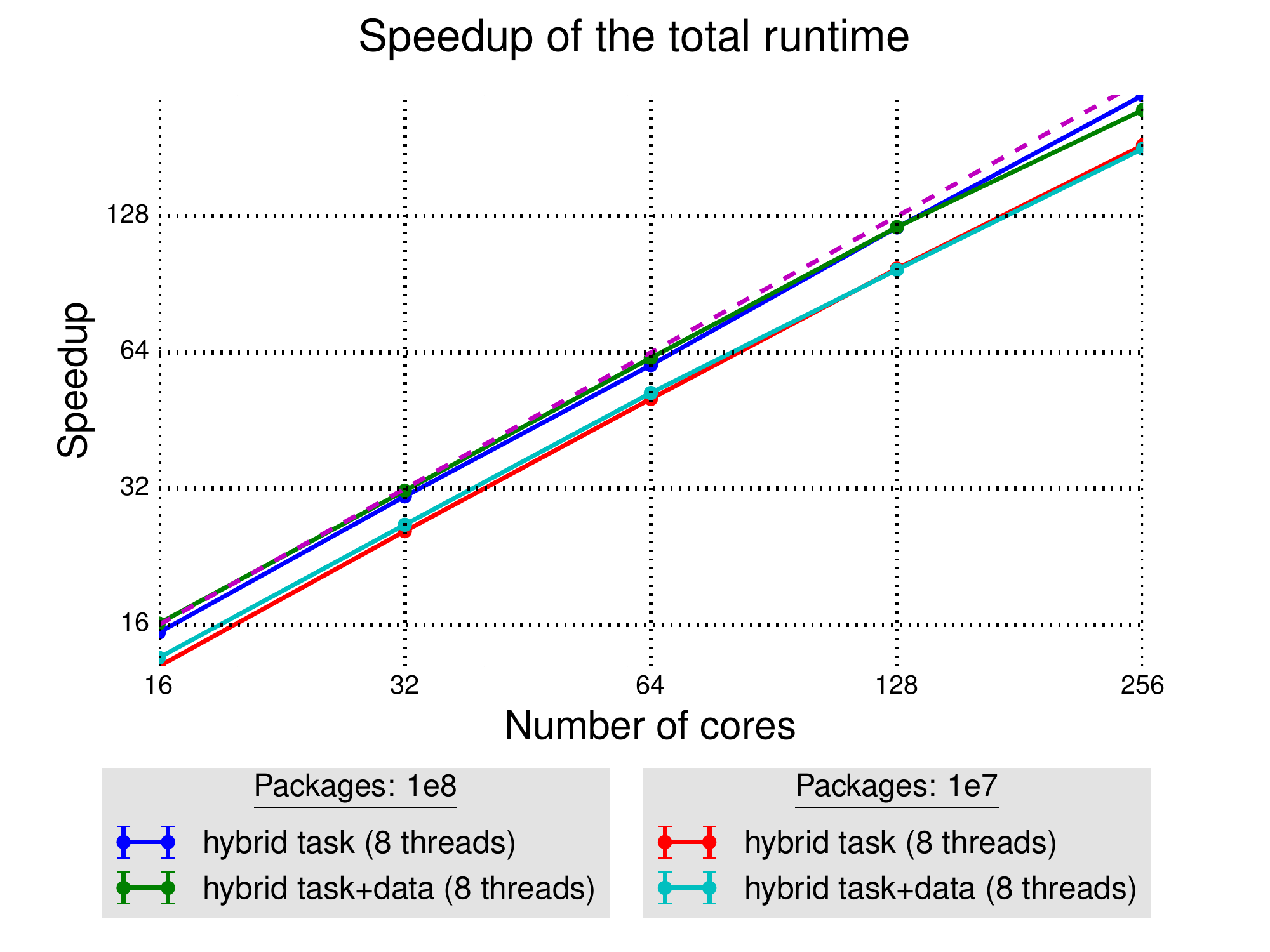}
\caption{Speedup of the simulation as a function of the number of cores, for $10^7$ and $10^8$ photon packages per wavelength, in both hybrid parallelization modes. The dashed curve represents the ideal speedup.}
\label{x10_speedups.fig}
\end{figure}

\begin{figure}[h!]
\centering
\includegraphics[width=0.5\textwidth]{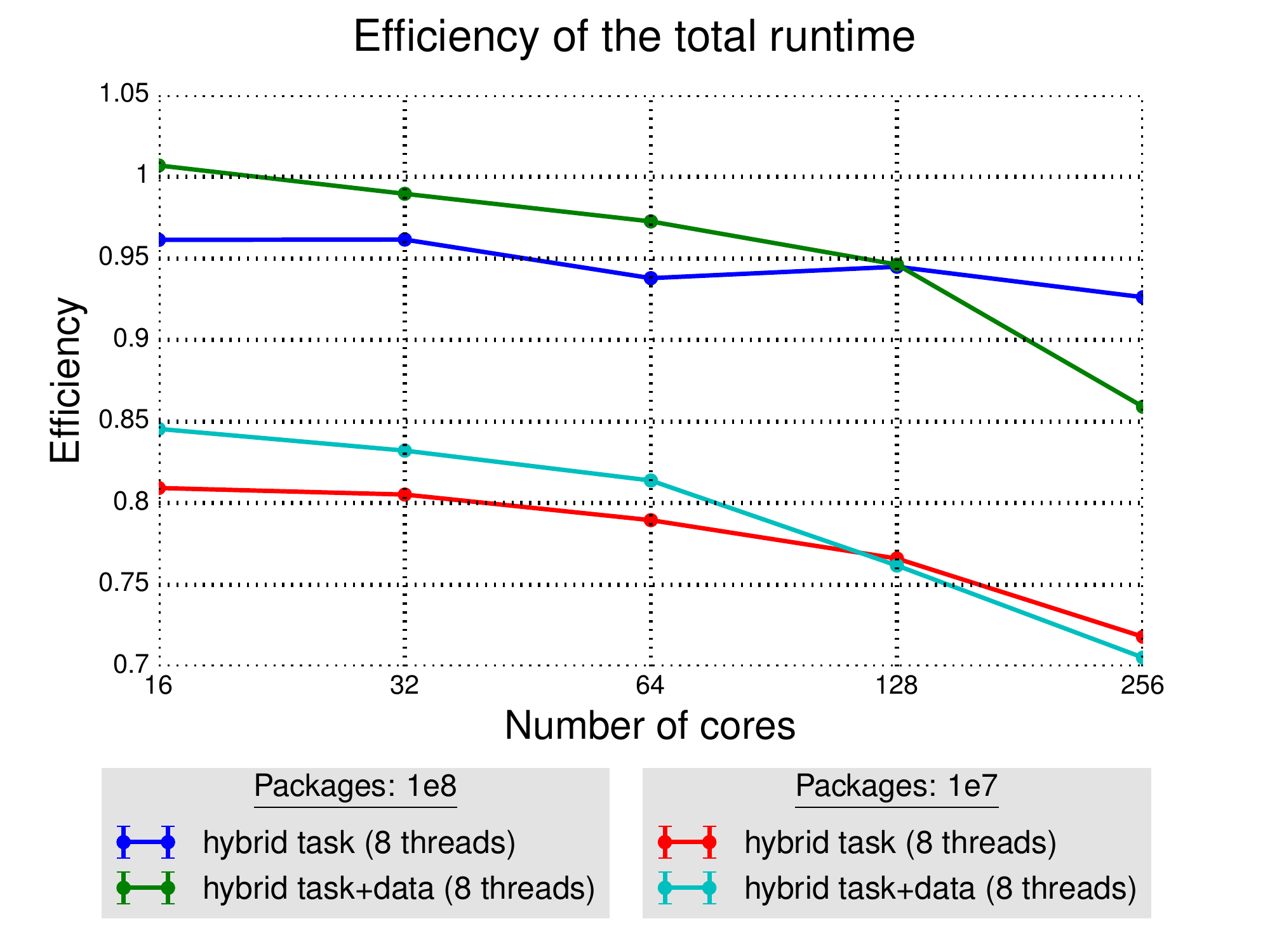}
\caption{Efficiency of the simulation as a function of the number of cores, for $10^7$ and $10^8$ photon packages per wavelength, in both hybrid parallelization modes.}
\label{x10_efficiencies.fig}
\end{figure}

To show how the scaling behaviour can depend strongly on the simulation parameters, we perform the same measurements with a number of photon packages increased by a factor of ten. The speedup and the efficiency are plotted in Figure \ref{x10_speedups.fig} and Figure \ref{x10_efficiencies.fig} respectively, for both the original and increased number of photon packages, and for both hybrid parallelization modes.


It is clear that the number of photon packages has an important impact on the parallel performance of the simulation. This is to be expected since increasing the number of photon packages directly gives more weight to the parallel portion of the algorithm relative to the serial part of the simulation. The best efficiency achieved for the simulation with $10^8$ photon packages per wavelength is 0.93, with a hybrid combination of 32 processes and 8 threads per process. Note that the efficiency of SKIRT can be easily raised even more when the number of photon packages (and/or wavelengths) is increased further.

\subsection{Memory scaling}
\vspace{0.5em}

\begin{figure}[h!]
\centering
\includegraphics[width=0.5\textwidth]{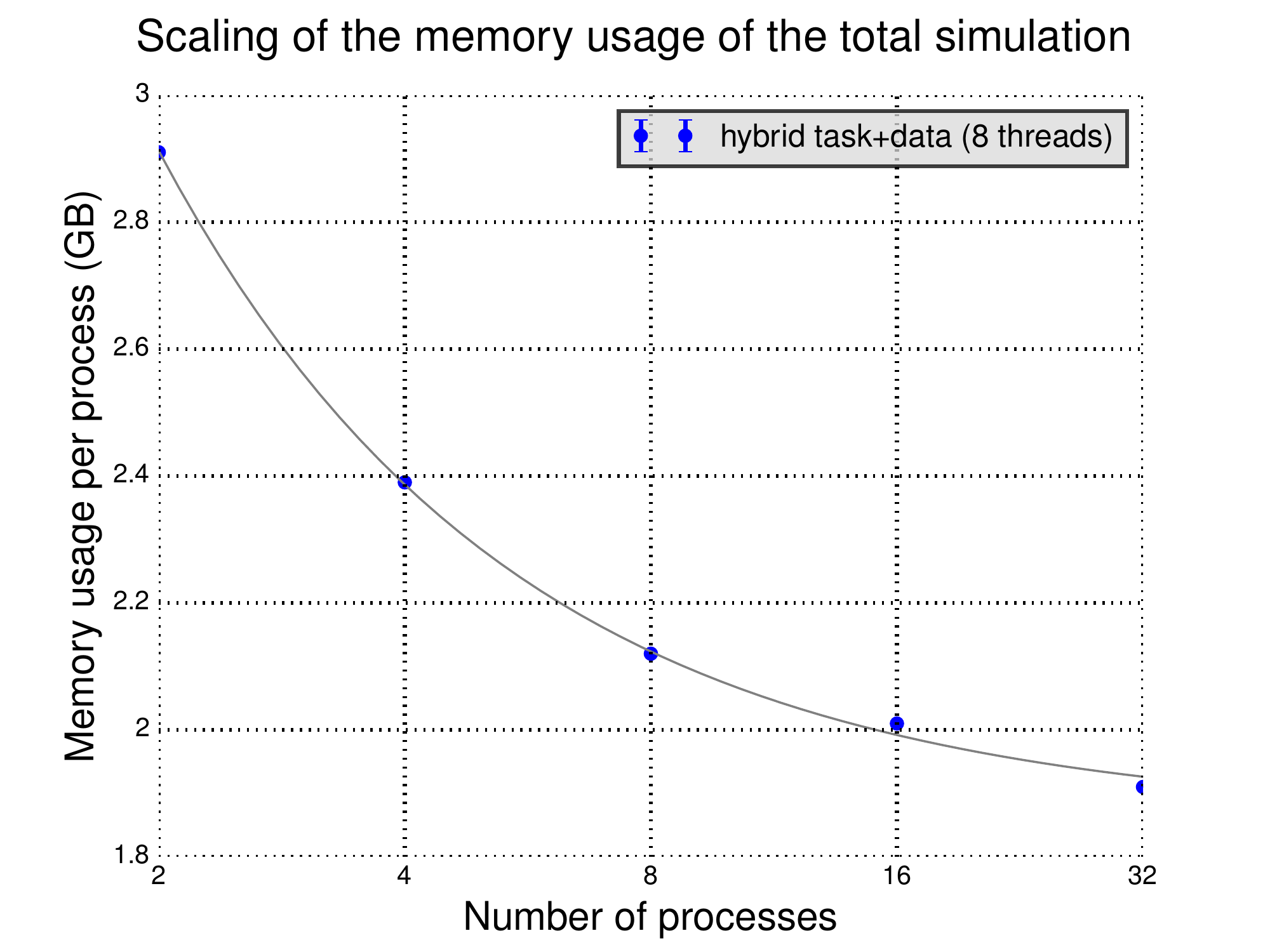}
\caption{The peak memory requirement of the simulation as a function of the number of processes. Two processes have been used per node, with eight threads per process. The grey line is a fitted curve of the form $y=a/x+b$. The best-fitting values for $a$ and $b$ were found to be 2.10 GB and 1.86 GB, respectively.}
\label{memoryscaling.fig}
\end{figure}

\noindent
As explained in subsection \ref{paralleltable.ssec}, in task+data parallel mode with $N_{\text{p}}$ processes, the memory footprint of each parallel table in SKIRT is reduced to $1/N_{\text{p}}$ of the non-parallel size. To test the impact of this, we record the peak memory requirement for the simulation for a varying number of processes. The result is plotted in Figure \ref{memoryscaling.fig}. 


\vspace{0.5em}
A function of the form $y=a/x+b$ has been fitted to the data points. Part of the data can be distributed amongst the processes and another part is `overhead' that needs to be allocated at each process. The first term in the equation represents the parallelizable part and the second term the constant part. The fitted relation with the parameters $a=2.10 \, \textrm{GB}$ and $b=1.86 \, \textrm{GB}$ shows that on one process, the parallelizable part corresponds to about 53\% of the simulation's memory requirements. This part shrinks by a factor of 32 when using 32 processes, corresponding to a decrease from 2.10 GB to just 66 MB. The total memory requirement per process has decreased to 49\% of its original value at 32 processes. The overhead causes the amount of liberated memory space to level off beyond 32 processes for this particular simulation. When more processes would be added, the \emph{global} memory consumption (from all processes combined) begins to increase linearly. However, it could be reasonable to do so because it may still induce a significant speedup (as long as resources are available).

\subsection{Simulation of a high-resolution galaxy model}
\vspace{0.5em}

\noindent
To acquire genuine memory and performance scaling results, simulations on a single core have to be run to normalize the recorded timings and memory measurements and infer the scaling behaviour. Therefore, tests such as those presented above are limited in resolution, to ensure that the serial runs still have a reasonable computation time and fit in the memory of the nodes.

\vspace{0.5em}
To demonstrate that the hybrid parallelization in SKIRT is not limited to these low-resolution models, we perform a simulation of the ERIS galaxy \citep{0004-637X-742-2-76}, a Milky Way-like galaxy model produced with a smoothed particle hydrodynamics (SPH) code.  
The ERIS galaxy has been used as a model in SKIRT in the past for an investigation of the dust energy balance in spiral galaxies using mock observations \citep{Saftly2015}. 

\vspace{0.5em}
For this particular simulation, three instruments have been configured, two with 5600 by 2000 pixels and one with 2048 by 2048 pixels. We use a logarithmic wavelength grid with 200 wavelength points. For the dust system, we use an octtree dust grid with a maximum of 13 division levels, resulting in a total of 8 172 431 dust cells. We run the simulation on the \mytexttt{Swalot} cluster on 10 nodes, with a total of 20 processes and 10 threads per process (2 processes per node). The number of photon packages used per wavelength is $10^8$, resulting in a total of 20 billion photon packages in the simulation.

\vspace{0.5em}
The total runtime of the simulation is 1d 8h 11m 2s (or 115862 seconds), corresponding to nearly 6500 CPU-hours. The peak memory usage per process is 26.3 GB so the peak memory usage per node is 52.6 GB, well below the available memory per node (128 GB). Without the hybrid task+data parallelization, this simulation would have been impossible to fit on such a node. Even on a system with enough memory, the simulation could be expected to run for at least two weeks.

\begin{figure*}[ht!]
\centering
\includegraphics[width=0.95\textwidth]{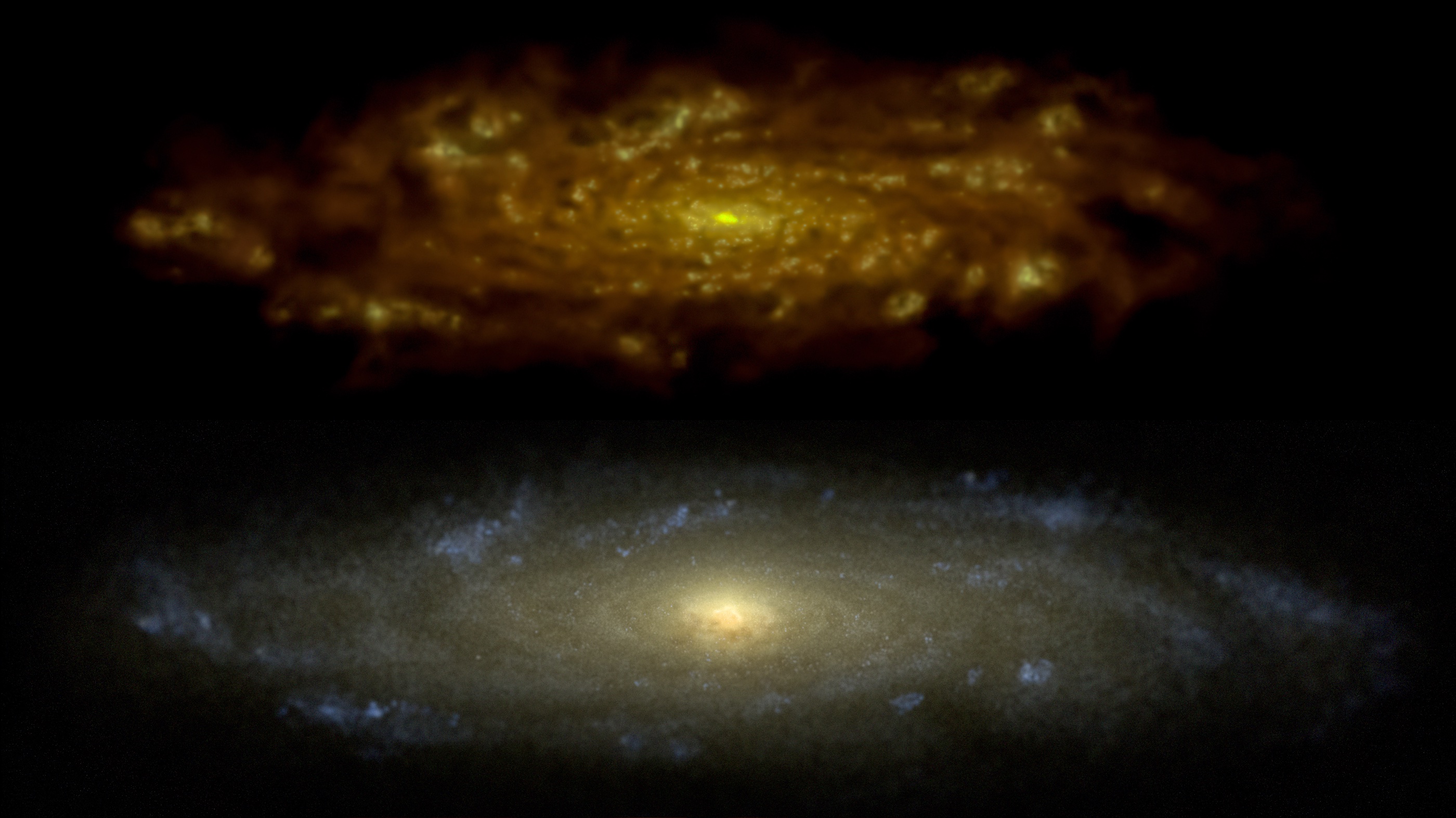}
\caption{High-resolution images produced from the output of a SKIRT simulation of the ERIS galaxy with an inclination angle of 75 degrees. Top panel: composite infrared image created from data at the wavelengths 33 micron, 55 micron and 77 micron. Bottom panel: optical RGB image created from the data at 0.33 micron, 0.55 micron and 0.77 micron}
\label{eris.fig}
\end{figure*}

\vspace{0.5em}
Figure \ref{eris.fig} shows two images generated from the output datacubes of the simulation. Both images show the galaxy at an inclination angle of 75 degrees. The top and bottom image are created from infrared and optical wavebands respectively.



\section{Conclusions}
\label{conclusions.sec}

\noindent
We have presented a new version of the dust radiative transfer code SKIRT, which is parallelized with a combination of shared and distributed memory programming. The task and data parallelization have been implemented with a strong modularity in design that enables a great flexibility towards different use cases and system architectures. Due to this flexibility and the use of lock-free programming, computing resources are used efficiently from single node cases to multi-node applications. Load imbalance is suppressed successfully by a combination of fixed task assignment for processes and dynamic task scheduling for threads. A strong scaling efficiency of 93\% at 256 cores is achieved for a simulation with 160 wavelengths and $10^8$ photon packages per wavelength. The efficiency depends strongly on the type of model and on the parallelization scheme, and increases with the number of photon packages and wavelengths. High-resolution simulations, particularly those with a large number of wavelengths, also benefit greatly from the implementation of a data parallelization scheme, where data structures are split across processes and the task assignment adjusted accordingly. This decreases the memory impact per process drastically. We designed abstractions for these data structures that offer a simple interface and can change representations internally when needed.

\section*{Acknowledgements}

\noindent
The computational resources (Stevin Supercomputer Infrastructure) and services used in this work were provided by the VSC (Flemish Supercomputer Center), funded by Ghent University, the Hercules Foundation and the Flemish Government ??? department EWI. This project has received funding from DustPedia\footnote{\hyperlink{http://www.dustpedia.com/}{http://www.dustpedia.com/}}, a European Union???s Seventh Framework Programme for research, technological development and demonstration under grant agreement no. 606874.

\section*{References}

\bibliography{references}

\end{document}